\documentclass[]{aa} 
\usepackage{graphicx}
\usepackage{txfonts}
\usepackage{color}
\usepackage{float}
\usepackage[normalem]{ulem}
\usepackage{natbib}
\usepackage{xcolor}
\usepackage[fleqn]{amsmath}
\usepackage{hyperref}
\hypersetup{colorlinks = true, linkcolor = {blue}, citecolor = {blue}, urlcolor= {blue}}
\linenumbers

\begin{document} 

\title{Celestial sunflowers}

\subtitle{Survival of rings around small planetary bodies under solar radiation pressure}

\author{
Zsolt Regály
    \inst{1,2}\thanks{E-mail: \href{mailto:regaly@konkoly.hu}{regaly@konkoly.hu}}\orcid{(0000-0001-5573-8190)}
\and
Viktória Fröhlich
    \inst{1,2,3}\orcid{0000-0003-3780-7185}
\and
Csaba Kiss
    \inst{1,2,3}\orcid{0000-0002-8722-6875}
}
\institute
{HUN-REN Konkoly Observatory, Research Centre for Astronomy and Earth Science, Konkoly-Thege Mikl\'os 15-17, 1121, Budapest, Hungary
\and
CSFK, MTA Centre of Excellence, Budapest, Konkoly Thege Miklós 15-17, H-1121, Budapest, Hungary
\and
ELTE E\"otv\"os Lor\'and University, Institute of Physics and Astronomy, P\'azm\'any P\'eter s\'et\'any 1/A, 1171 Budapest, Hungary}

\date{Received \today; peer review in progress}
 
\abstract
{
Rings around giant planets are a common feature of the Solar System.
Even though solar radiation pressure is known to destabilize rings by exciting the orbital eccentricity of its particles, the Centaur Chariklo (and possibly Chiron), the dwarf planet Haumea, and trans-Neptunian object Quaoar also host rings of solid material.
}
{
We explore the dynamical evolution of rings around spherical Chariklo and Haumea analogs, assuming different particle sizes and tilt angles with respect to the planetary orbital plane of the ring.
}
{
The ring dynamics were studied using a GPU-based N-body integrator with an 8th-order Hermite scheme for several thousand years, corresponding to 10 solar orbits.
The simulations took into account the gravitational effects of the planet and the Sun, radiation pressure, and the shadow cast by the planet.
}
{
Two families of rings have been identified depending on the ring tilt angle.
Slightly tilted rings ($\leq40^\circ$) are unstable under a critical particle size.
Highly tilted rings ($\geq50^\circ$), however, show instability only for a range of particle sizes that spans 1–10 times the critical size.
The planetary shadow reduces the critical size by a factor of five and extends the instability region to 0.1–10 times this newly identified critical size.
}
{
The stabilization of highly inclined rings occurs because the plane of the ring is forced to be perpendicular to the Solar radiation.
As a result, the plane of the ring rotates as the planetary bodies revolves: always facing the sun, like a celestial sunflower.
Rings which are closely aligned to the orbital plane of the host planet, such as Haumea and Quaoar, presumably consist of particles with a size at least $1-4~\mu$m.
However, particles in the rings which are highly tilted, like that around Chariklo and Chiron, should consist of particles $\lesssim2.5-15~\mu$m or $\gtrsim60-300~\mu$m
}


\maketitle

\section{Introduction}

Ring systems, also known as planetary rings, are a common component of our Solar System.
Rings are composed of solid material, including dust, meteoroids, planetoids, and moonlets.
In the Solar System, all of the giant planets are observed to possess ring systems.
Nevertheless, planets within the orbit of Jupiter do not possess such a ring system.
Today, 5042 Trans Neptunian Objects (TNOs) and Centaurs are known to orbit beyond Neptune and between Jupiter and Neptune\footnote{https://www.minorplanetcenter.net}.
So far four ring systems have been discovered around small Solar System bodies by observing stellar occultations. 
These include Chariklo \citep{Braga-Ribasetal2014Natur.508...72B}, Chiron \citep{Ortizetal2015A&A...576A..18O}, Haumea \citep{Ortizetal2017Natur.550..219O}, and Quaoar \citep{Morgadoetal2023Natur.614..239M,Pereiraetal2023A&A...673L...4P}.
The occurrence of rings around small planetary bodies is estimated to be relatively common \citep[8-12 percent,][]{Sicardyetal2020tnss.book..249S,Sicardy2025RSPTA.38340193S}.

Chariklo, the first non-giant planet discovered to host rings, possesses two dense and narrow rings located at 391~km and 405~km from its center, composed partly of water ice \citep{Braga-Ribasetal2014Natur.508...72B}.
Chiron may have a broad $\sim$580 km disk with concentrations at radii of 325$\pm$16 km and 423$\pm$11 km  \citep{Ortizetal2023A&A...676L..12O}.
The ring of Chariklo is tilted with respect to the orbital plane, as is the putative ring of Chiron.
Haumea's ring, situated near the 3:1 spin-orbit resonance, is about 70~km wide and is coplanar with its equator \citep{Ortizetal2017Natur.550..219O}.
Quaoar's ring, an unexpected discovery, challenges the Roche limit theory as it lies well beyond the theoretical boundary of the Roche radius \citep{Morgadoetal2023Natur.614..239M}.
The latter two rings are close to the orbital plane of their respective planetary body.
These discoveries highlight the diverse mechanisms that can lead to the formation and stability of rings around smaller celestial bodies.

The rings around small Solar System objects consist of particles assumed to be varying in size from fine dust to large debris.
Chariklo's rings are composed of particles ranging from millimeters to meters, with optical depths similar to Saturn's A ring, indicating significant collisional activity and a wide particle size distribution \citep{Braga-Ribasetal2014Natur.508...72B}.
For Chiron, suspected ring material may include icy particles ranging from sub-millimeter to meter sizes, though precise measurements remain elusive \citep{Ortizetal2015A&A...576A..18O}.
Haumea's ring particles are likely larger than a few microns, with sizes extending to the centimeter to meter range, as inferred from its consistent opacity across wavelengths \citep{Ortizetal2017Natur.550..219O}.
Quaoar's rings, located beyond the Roche limit, likely consist of particles spanning micrometer to meter in size, shaped by gravitational and collisional dynamics \citep{Sicardyetal2020tnss.book..249S}.
These particle sizes reflect the dynamic processes governing ring formation and stability in diverse environments.

The seminal work by \citet{Burns1979} provided a unifying theoretical framework that remains pivotal for understanding how solar radiation affects the dynamics of planet-orbiting particles.
By introducing the $\beta$-parameter which quantifies the ratio of radiation pressure to gravitational attraction and describing the role of the Poynting–Robertson drag \citep{Wyatt1950ApJ...111..134W} and the Yarkovsky effect \citep{Hasegawa1977MmKyo..35..131H}, their study elucidated how micron- to millimeter-sized particles shape both planetary ring systems and interplanetary dust populations.
Applying analytical solutions to the perturbation equations, the authors showed that particles of rings formed by capturing interplanetary debris, scattering of lunar ejecta, or randomly ejected materials are subject to gain eccentricity due to radiation pressure.
As a result, particles can collide with or leave the Hill sphere of the central object. 

\cite{Mignard1982Icar...49..347M} extended the study of \citet{Burnsetal1979Icar...40....1B} by deriving differential equations that consider the dynamics of small particles orbiting the planet in inclined orbits.
Using numerical integrations he found that for small ring inclinations ($\lesssim10^\circ$) the coplanar solution is valid, but for large inclination angles ($\gtrsim75^\circ$) the ring plane shows strong vibrations and reduced eccentricity excitation due to radiation pressure.
Subsequent extensions refined this model by incorporating more detailed scattering properties \citep{Mukai1989}, nonspherical particle shapes \citep{Kimura1997}, and additional torques such as the Yarkovsky–O’Keefe–Radzievskii–Paddack (YORP) effect \citep{Rubincam2000}. 

The dynamics of the rings around the giant planets of the Solar System have been studied extensively in the past.
The Jupiter ring system under the influence of the Galilean moons, the planetary magnetic field, and solar radiation has been studied by several authors (see, e.g., \citealp{Krivov2002JGRE..107.5002K,Sachse2018Icar..303..166S,Kane2022LPICo2687.3041K}).
As a general conclusion, the Galilean moons and their associated resonances can strongly influence the long-term stability of the rings.
It is known that Saturn's C, B, and inner A rings are depleted of small particles in the cm-dm range \citep{French2000Icar..145..502F,Thomson2005AGUFM.P33B0247T}, likely due to radiation pressure efficiently removing cm-sized particles \citet{Rubincam2006Icar..184..532R}.
In a study that includes the effect of the planetary shadow, \citet{Vokrouhlicky2007A&A...471..717V} have shown that the Poynting-Robertson drag dominates the dynamics for small particles, which can drift inward throughout the ring system.
In such exoplanet systems, where the ring-host planet is close to the central star, the plane of the tilting rings above a critical tilt angle exhibits oscillations due to the Lidov-Kozai mechanism, which may lead to strange light curve features \citep{Sucerquia2017MNRAS.472L.120S}.

The $\mu$ and $\nu$ rings of Uranus are radially broad and are dominated by micrometer-sized dust \citep{dePater2006Sci...312...92D,dePater2006Icar..180..186D,Showalter2006Sci...311..973S}.
Numerical simulations by \citealp{Sfair2009A&A...505..845S} have shown that the Poynting-Robertson drag causes the collapse of the orbits on a timescale of $3.1\times10^5-3.6\times10^6$ years. 
Radiation pressure excites, while the oblateness of Uranus tends to decrease the eccentricity of the particles.
Encounters with satellites further complicate the evolution of the particle orbits.
The Neptunian ring system is also populated by micron-sized dust \citep{Smith1986Sci...233...43S,Smith1989Sci...246.1422S}.
\citep{Ferrari1994Icar..111..193F} have shown that the four arcs and clumps in the Adams ring contain large particles embedded in the dust.
Recent N-body simulations by \citet{Madeira2022MNRAS.510.1450M} suggest that these arcs can be  partially explained by the dynamics of co-orbital moonlets.
Rings around moons of giant planets seem to be stable configurations \citep{Sucerquia2024A&A...691A..74S}, but their absence in the Solar System can be attributed to non-gravitational phenomena such as stellar radiation, magnetic fields, and the influence of magnetospheric plasma.

The dense and narrow rings around small Solar System bodies have been studied to explain why their rings are relatively far away from the central body when scaled to those of the giant planets. 
It has been shown that the gravitational fields of small bodies such as Chariklo and Haumeea have large non-axisymmetric terms that create strong resonances between the spin of the object and the mean motion of the ring particles \citep{Sicardy2020AJ....159..102S,Sumida2020ApJ...897...21S}.
Recent N-body simulations suggest that Chariklo's ring system could be dominated by several satellites that shape the ring edges \citep{Winter2023A&A...679A..62G} or by a single satellite with a mass of a few $\times10^{13}$~kg that is in orbital resonance with the rings \citep{Sickafoose2024PSJ.....5...32S}.
We emphasize that the effect of solar radiation or ring tilt angles has not been investigated in detail in these studies of small Solar System bodies.

To extend our knowledge on rings around small Solar System bodies, we present the results of our investigation of the effect of solar radiation pressure on the dynamics and lifetime of rings, assuming different ring tilt angles, particle sizes and heliocentric distances.
The simulations are carried out using a GPU-based high-precision N-body integrator that incorporates the effects of radiation pressure.
Given the proximity of the ring to the planetary body, the effect of its shadow was also considered.

The paper is organized as follows.
Section~2 presents an analytical model that predicts the critical $\beta$ parameter for ring stability assuming that the ring plane is in the orbital plane of the planetary body.
Section~3 deals with the limitations of the analytical model, such as the planetary shadow and the ring tilt angle, using a numerical N-body model.
The results of the numerical simulations are presented in Section~4.
This is followed by a discussion in Section~5.
The paper closes with our conclusions in Section~6.

\section{Analytical model}

\begin{figure}
    \centering
    \includegraphics[width=1\linewidth]{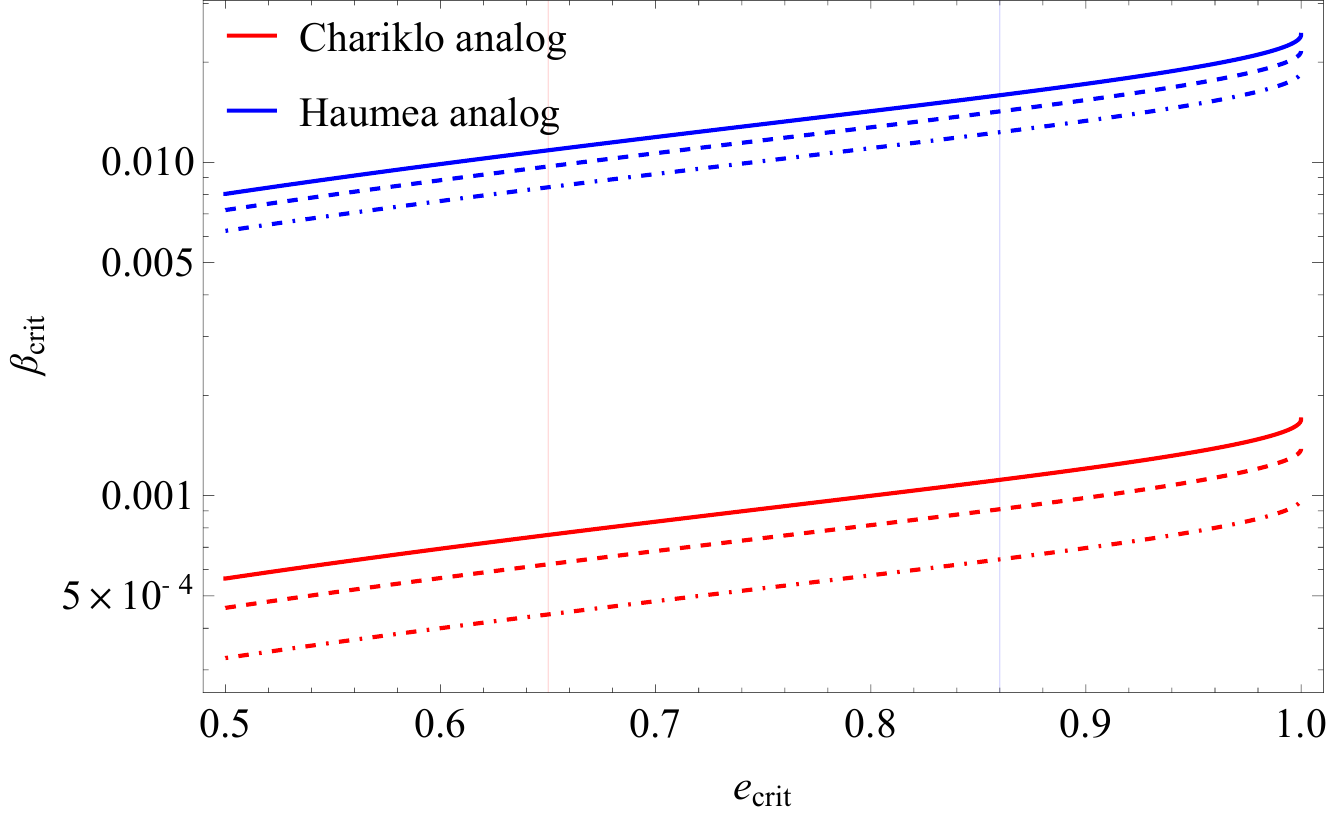}
    \caption{$\beta_\mathrm{crit}$ as a function of $e_\mathrm{crit}$ for the Chariklo and Haumea analogs. 
    Three different semimajor axes are assumed for the planetary bodies (dot dashed, dashed, and solid lines represent $a_\mathrm{p}=5,~10,~15$au for the Chariklo and $a_\mathrm{p}=30,~40,~50$ au for the Haumea analog.
    Vertical lines show the critical eccentricities in the two ring systems.}
    \label{fig:beta-ecc}
\end{figure}

The normalized radiation pressure force, $\beta$, acting on a particle orbiting the Sun is
\begin{equation}
    \beta = \frac{|\textbf{F}_{\mathrm{rad}}|}{|\textbf{F}_{\mathrm{grav}}|},
    \label{eq:beta}
\end{equation}
where $\textbf{F}_{\mathrm{grav}}$ and $\textbf{F}_\mathrm{rad}$ are the Sun's gravitational and radiation forces acting on a given ring particle.
Since both forces are inversely proportional to the distance squared, $\beta$ depends only on the properties of the particle and the radiation source, see the next section for details.
Particles orbiting a star remain on a bound elliptical orbit for $\beta < 0.5$, but for $\beta > 0.5$ the particles' orbit becomes parabolic or hyperbolic, causing them to leave the central star \citep{Wyatt1999ApJ...527..918W}.

However, the situation changes slightly if the particles are orbiting a small planetary body, because in this case, even if $\beta>0.5$, their orbit will not become hyperbolic. 
This is because the particles are in the potential well of the planetary body.
Escaping from planetary orbit requires that particles are beyond the planet's Hill sphere at apocenter.
Moreover, eccentric particles will remain in orbit as long as their pericenter distance remains greater than the radius of the planetary body.
Above a critical eccentricity, 
\begin{equation}
     e_\mathrm{crit}=1-(D_\mathrm{p}/2)/R_0
     \label {eq:ecrit}
\end{equation}
the particles hit the surface at the pericenter distance, where $D_\mathrm{p}$ is the diameter of the planetary body and $R_0$ is the semimajor axis of the particles' orbit.

\citet{Burnsetal1979Icar...40....1B} presented a solution to the perturbation equation of celestial mechanics using a method developed by \citet{Herrick1948PASP...60..321H}, where the perturbations are averaged over the parent planet's orbital period.
The solutions revealed that the eccentricity of a particle in a ring oscillates between 0 and $e_\mathrm{max}$ with the orbital period of the  planetary body.
By assuming zero inclination for the orbital planet of the ring with respect to that of the planetary body, the maximum eccentricity of the particles located at a distance of $R_0$ around a planet with mass $M_\mathrm{p}$ orbiting the Sun at a semimajor axis of $a_\mathrm{p}$ is
\begin{equation}
    e_\mathrm{max}< \sin\left(\frac{3}{2}\pi\beta\sqrt{\frac{M_\odot}{M_\mathrm{p}}\frac{R_0}{a_\mathrm{p}}}\right).
    \label{eq:e_crit}
\end{equation}
If $e_\mathrm{max}=e_\mathrm{crit}$ for which case the particles will hit the planetary body's surface a critical $\beta$ can be given as 
\begin{equation}
    \beta_\mathrm{crit}>\frac{2}{3}\frac{1}{\pi}\sqrt{\frac{M_\mathrm{p}}{M_\odot
}\frac{a_\mathrm{p}}{R_0}}\arcsin\left(e_\mathrm{crit}\right).
    \label{eq:beta_crit}
\end{equation}
We note that the above expression is not equivalent to the $\beta_\mathrm{crit}$ defined in \citet{Burnsetal1979Icar...40....1B}, since that corresponds to $e_\mathrm{max}=1$.
Figure~\ref{fig:beta-ecc} shows the critical $\beta$ as function of the critical eccentricity.
In this study, two small body ring systems are modeled: spherical Chariklo and a Haumea analogs.
According to Eq.~(\ref{eq:e_crit}), the critical eccentricities for the Chariklo and Haumea analogs are $e_\mathrm{crit}=0.65$ and $0.83$, respectively , assuming the diameters and ring distances shown in Table~\ref{tab:physparam}.
Thus, the $\beta_\mathrm{crit}$ parameters above which the rings quickly become unstable are $5\times10^{-4}$ and $10^{-2}$ for the Chariklo and Haumea analogs, respectively.

In the above analytical model several assumptions are made: 1) sunlight reflected from the planet's surface is neglected \citep{Shapiro1963}; 2) Solar flux is assumed to be constant (i.e., a circular orbit is assumed); 3) the shadow cast by the planets is not taken into account \citep{Allan1962SatelliteOP,Radzievskii1962SvA.....5..758R}; 4) interactions with the planetary magnetic field are not taken into account; 5) the planet is assumed to be spherical, meaning that resonance effects\footnote{We note that all known Centaur and TNO rings are close to 3:1 spin-orbit resonance.} between, for example, orbital precession rates and the motion of the planetary ring particles are not included (see \citealp{Shapiro1963} and \citealp{Allan1967}); 6) interactions with planetary satellites are ignored.   
To address some of the above simplifications (effect of high ring tilt angles and the planetary shadow), we performed an extensive numerical experiment, which is detailed in the next section.

\section{Numerical model}

The effect of radiation pressure on the ring particles was modeled using an 8th order high-precision GPU-assisted Hermite N-body integrator (see, e.g., \citealp{NitadoriMakino2008NewA...13..498N,Regalyetal2018MNRAS.473.3547R,DencsRegaly2019MNRAS.487.2191D,DencsRegaly2021A&A...645A..65D}). 
The numerical integration took into account both the radiation pressure exerted by the Sun and the shadow cast by the planetary body on the ring. 
The acceleration of a ring particle is 
\begin{equation}
	\ddot{\textbf{R}_i}=-G \left[M_\mathrm{p}\left(\frac{\textbf{r}_i}{r_i^3}\right)-M_\odot(1-\gamma_\mathrm{shadow}\beta)\left(\frac{\textbf{R}_i}{R_i^3}\right)\right],
    \label{eq:motion}
\end{equation}
where $M_\mathrm{p}$ and $M_\odot$ are the mass of the planetary body and Sun, respectively. $\textbf{r}_i$ and $\textbf{R}_i$ represent the $i^\mathrm{th}$ ring particle's distance from the planetary body and Sun, respectively. 
$\gamma_\mathrm{shadow}=0~|~1$ specifies whether the particle is in the shadow of the planetary body or is illuminated by the Sun, respectively.
A thorough description of the algorithm that determines the value of the shadow parameter is provided in Appendix~\ref{apx:shadow}.
In this study, the Poynting-Robertson drag has been neglected because its effect is negligible for $\beta\leq0.8$ compared to that of the radiation pressure on the investigated timescale, see details in Sect.~\ref{sec:caveats}.
The Yarkovsky effect -- force arises due to asymmetric absorption and reradiation of the received solar energy for a rotating
body having thermal lags -- has also been neglected because it dominates other dissipative perturbations for bodies that are meter-sized bodies \citet{Burns1979}.

The radiation pressure force according to \cite{Burns1979} can be given as
\begin{equation}
    |\textbf{F}_{\mathrm{rad},i}| = S_\odot \left(\frac{R_i}{1 \mathrm{au}}\right)^{-2} \pi s^2 c^{-1} \left<Q_{pr}(s)\right>,
    \label{eq:Frad}
\end{equation}
where $S_\odot=1.367$~kW~m$^{-2}$ is the solar constant, $R_i$ is the heliocentric distance of the given particles in (au), $s$ is the particle radius and $c$ is the speed of light.
In Eq.~(\ref{eq:Frad}) $\left<Q_\mathrm{pr}(s)\right>$ is the average radiation pressure efficiency over the stellar spectrum acting on a grain with radius $s$.
$\left<Q_{pr}(s)\right>$ is calculated using the optical constants of the grains, which depends on the composition and geometrical structure of the grain.
Assuming that $|\textbf{F}_{\mathrm{grav},i}|=-GM_\odot/R_\mathrm{i}^2$, the normalized radiation pressure force given by Eq.~(\ref{eq:beta}) can be given as
\begin{equation}
    \beta=\frac{3}{4c G}\left(\frac{S_\odot}{M_\odot}\right)\left(\frac{\left<Q_{pr}(s)\right>}{\rho s}\right),
    \label{eq:beta-Qprs}
\end{equation}
where $G$ is the Newtonian gravitational constant and $\rho$ is the intrinsic density of the particle.
Details of the calculation of $\left<Q_\mathrm{pr}(s)\right>$ and $\beta$ are given in  the Appendix~\ref{apx:beta-Mie}.

To solve Eq.~(\ref{eq:motion}), we used an adaptive time-step for the integrator recently tested by \citet{Phametal2024OJAp....7E...1P} in the form of
\begin{equation}
    \Delta t=\mathrm{min}\left[\left(2 \frac{\textbf{R}_i^{(2)}\textbf{R}_i^{(2)}}{\textbf{R}_i^{(3)}\textbf{R}_i^{(3)}+\textbf{R}_i^{(2)}\textbf{R}_i^{(4)}}\right)^{1/2}\right],
\end{equation}
where $^{(2)}, ^{(3)}$ and $^{(4)}$ represent the second, third, and fourth order derivatives, respectively.

We modeled the ring with 132,000 particles which were initially in a circular orbit around the planetary body.
If the particle reached the surface of the cental body, meaning their distance reached the body's radius, it was removed from the simulation.
The simulation was stopped if only 0.1 percent of the particles remained in the system or if the planetary body completed 10 orbits around the Sun.
The elapsed time in the simulations can be referred to as the ring's lifetime.
We considered the ring to be stable on a timescale of a few millennia if the lifetime of the ring exceeded 10 solar orbital periods.

We have assumed that the planetary body is a spherical body, orbiting the Sun in a circular orbit, and has no companion moon(s).
For the planetary body, we considered two different masses, $M_\mathrm{p}$, that of Chariklo and that of Haumea.
We investigated three different solar orbital semimajor axes, $a_\mathrm{p}$, for the planetary bodies in the ranges of $5~\mathrm{au}-15~\mathrm{au}$ and $30~\mathrm{au}-50~\mathrm{au}$ for the Chariklo and Haumea analogs, respectively.
The ring widths, $\Delta R$, are assumed to be those derived from observations.
The physical parameters of planetary body analogs are summarized in Table~\ref{tab:physparam}.

Regarding the initial tilt angle of the ring plane with respect to the orbital plane of the planetary body, ten different values were investigated in the range of $0^{\circ}\leq i \leq 90^{\circ}$.
The size of the grains in the rings of known Solar System small bodies (Chariklo, Chiron, Haumea and Quaaoar) usually ranges from submicron to millimeter scales.
Thus, to represent these particles, 11 different $\beta$ values were assumed with a logarithmical distribution in the range of $7.8125\times10^{-4}-0.8$.
The rings consisted of 132,069 massless particles distributed homogeneously across a 10~km and 70~km wide region for the Chariklo and Haumea analogs, respectively.
The inner ring radius was set to match that of the discovered rings around Chariklo and Haumea.
Particles were initially in circular orbits around the planetary body.
The investigated parameters of rings are summarized in Table~\ref{tab:paramspace}.

\begin{table}
   \caption{Physical parameters of small planetary bodies and rings used in the numerical simulations.}
    \centering
    \begin{tabular}{c|c|c}
        \hline\hline
        Parameter & Chariklo analog\tablefootmark{a} & Haumea analog\tablefootmark{b}\\
        \hline
        $M_\mathrm{p}$ (kg)         & $8.82\times10^{18}$ & $4.0\times10^{21}$ \\
        $D_\mathrm{p}$ (km)         & $256.32$ & $780$ \\
        $R_0$ (km)                  &  $386$ & $2287$ \\
        $\Delta R$ (km)                  & $10$ & $70$\\
        $a_\mathrm{p}$ (au)                    & 5, 10, 15 & 30, 40, 50 \\
        \hline
    \end{tabular}
    \label{tab:physparam}
    \tablefoot{Parameters are taken from:
    \tablefoottext{a}{\citet{Kondratyev2016Ap&SS.361..389K}} and 
    \tablefoottext{b}{\citet{Dunham2019ApJ...877...41D}}
    }
\end{table}

\begin{table}
   \caption{Investigated ring parameters in the numerical simulations.}
    \centering
    \begin{tabular}{l|l}
        \hline\hline
        Parameter & Values \\
        \hline
        $i$ (deg)                   & 0, 10, 20, 30, 40, 50, 60, 70, 80, 90 \\
        $\beta$                     & $7.8125\times10^{-4},~1.5625\times10^{-3},~3.125\times10^{-3},$ \\
                                    & $6.25\times10^{-3},~1.25\times10^{-2},~2.5\times10^{-2},$ \\
                                    & $5\times10^{-2},~0.1,~0.2,~0.4,~0.8$ \\
        \hline
    \end{tabular}
    \label{tab:paramspace}
\end{table}

\section{Results}

\begin{figure*}[h!]
    \centering
    \includegraphics[width=0.9\linewidth]{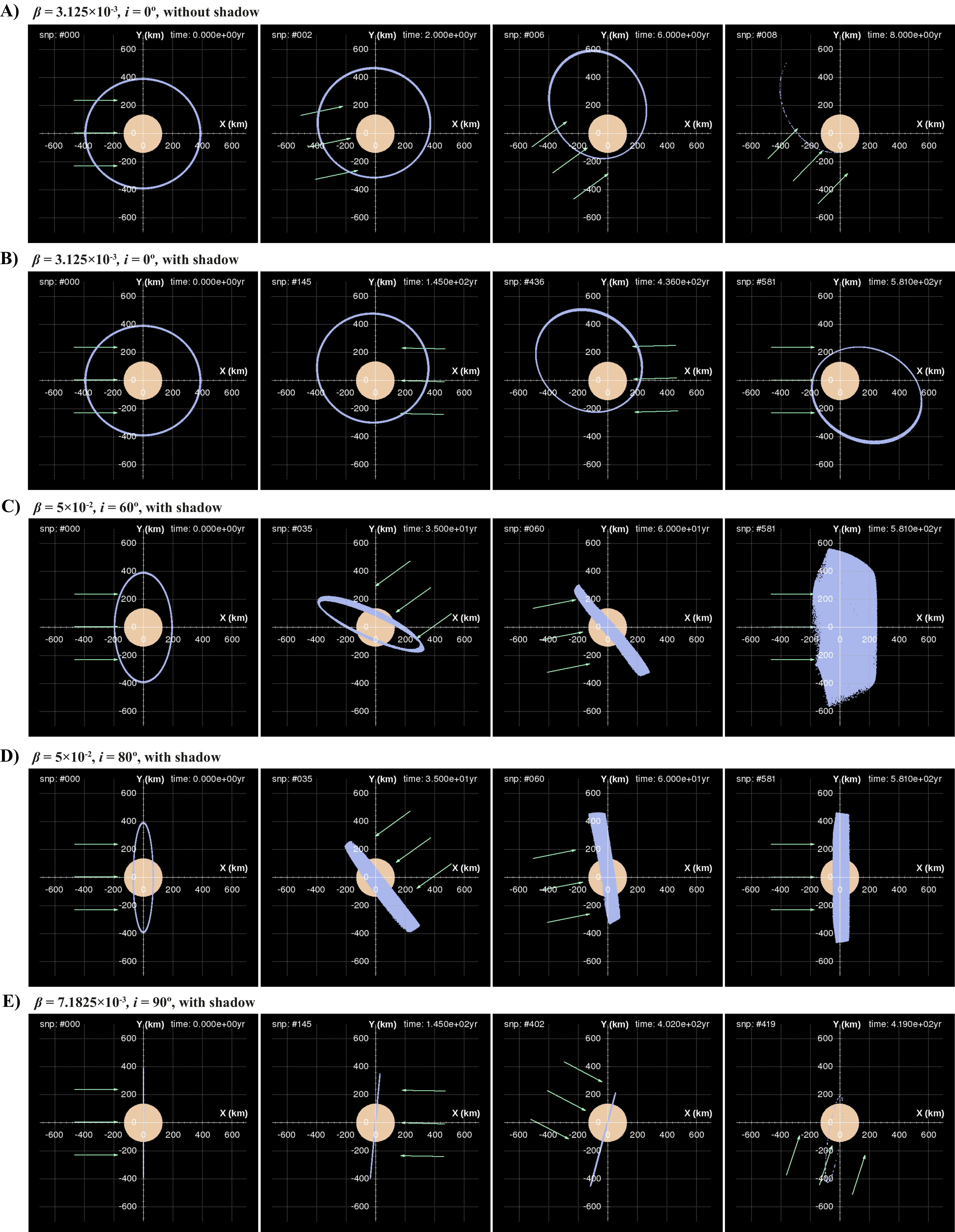}
    \caption{Some examples of how the ring has changed in Chariko analogue models seen from above. 
     The direction of the radiation that originates from the Sun is indicated by green arrows.
    Panel A: Unstable coplanar ring model (ring lifetime is about 0.1 orbital period of the planetary body).
    Panel B: Stable coplanar ring model (particles do not reach the critical eccentricity at the end of the simulation).
    Panel C and D: Stable ring models with initial inclination angles of $50^\circ$ and $80^\circ$, respectively.
    Panel E: Unstable ring model with an initial tilt angle of $90^\circ$.
    Movies that were generated from the simulation snapshots can be found online.
    }
    \label{fig:snapshots}
\end{figure*}

\subsection{Overview of ring dynamics}

First let's examine models in which the planetary shadow on the rings' particles is neglected.
As it is predicted by the analytical model, coplanar rings ($i=0^\circ$) become unstable above a critical $\beta$ given by Eq.~(\ref{eq:beta_crit}).
Panel A of Fig.~\ref{fig:snapshots} shows four snapshots of the ring evolution for a Chariklo analog, where $\beta=3.125\times10^{-3}$, which is greater than $\beta_\mathrm{crit}$.
As can be seen, the ring particles collide with the planetary body when particle eccentricity reaches a critical value ($e_\mathrm{crit}=0.65$).
This happens very early, at about 0.1 solar orbital period.

According to \citet{Burnsetal1979Icar...40....1B}, the effect of the planetary shadow is negligible.
As demonstrated by \citet{Allan1962SatelliteOP} and \citet{Radzievskii1962SvA.....5..758R}, the character of the perturbation of particle orbits are unaffected by the shadow; only its magnitude changes by about ten percent.
Therefore, studies of the dynamics of small dust particle orbiting a Solar System planet usually neglect the effect of planetary shadow (see, e.g., \citealp{Mignard1982Icar...49..347M, HamiltonKrivov1996Icar..123..503H, KovacsRegaly2018MNRAS.479.4560K}).
However, if the shadow is taken into account, the eccentricity of the particles cannot grow beyond the critical value due to the weaker perturbation caused by the average reduced irradiation time. 
When the pericenter of the particle is in shadow, this effect is particularly strong. 
Thus, if $\beta$ is close to $\beta_\mathrm{crit}$, the approximately ten percent change in magnitude of the radiation effect due to shadowing can be enough to reduce the effective $\beta$ below the critical value, allowing particles to remain stable.
As a result, the ring can be stable on the investigated timescale, as is shown in panel B of Figure \ref{fig:snapshots}.

For a sufficiently distant ring, it may happen that even below the critical eccentricity, the apocenter distance exceeds the Hill radius, allowing the particle to escape.
However, in the case of the rings of the studied planetary bodies, this never occurs because the particle impacts the planet's surface at pericenter first.
We note that the particles' eccentricity oscillates throughout the simulation as pointed out by \citet{Burnsetal1979Icar...40....1B}, see details in Sect.~\ref{sec:Disc}.
The above observations maintain their validity across the full range of investigated orbit distances and planetary masses for initially tilted rings with inclination angles of $i\leq40^\circ$.

Nevertheless, a wholly novel phenomenon emerges for higher initial ring tilt angles, $i\geq50^\circ$.
For relatively high values of $\beta$, the ring can be stabilized (see panels C and D in Fig.~\ref{fig:snapshots}, that show snapshots of two models assuming initial ring tilt angles of $i=50^{\circ}$ and $i=80^{\circ}$, respectively, for the same $\beta=5\times10^{-2}$).
As one can see, the rings survive until the end of the simulation and their tilt angle becomes perpendicular to the direction of radiation pressure.
Another phenomenon is that the initially highly tilted ($i\geq50^\circ$) rings become unstable below the critical value of $\beta_\mathrm{crit}$, for which case the slightly tilted ($i\leq40^\circ$) or coplanar rings are found to be stable. 
We note that the lowest value of $\beta$ studied is still insufficient to maintain a $i=90^\circ$ stable ring around a Chariklo analog, see panel E of Fig.~\ref{fig:snapshots}.
However, assuming a higher planetary mass, the ring becomes stable, see next section. 

\begin{figure*}
    \centering
    \includegraphics[width=1\linewidth]{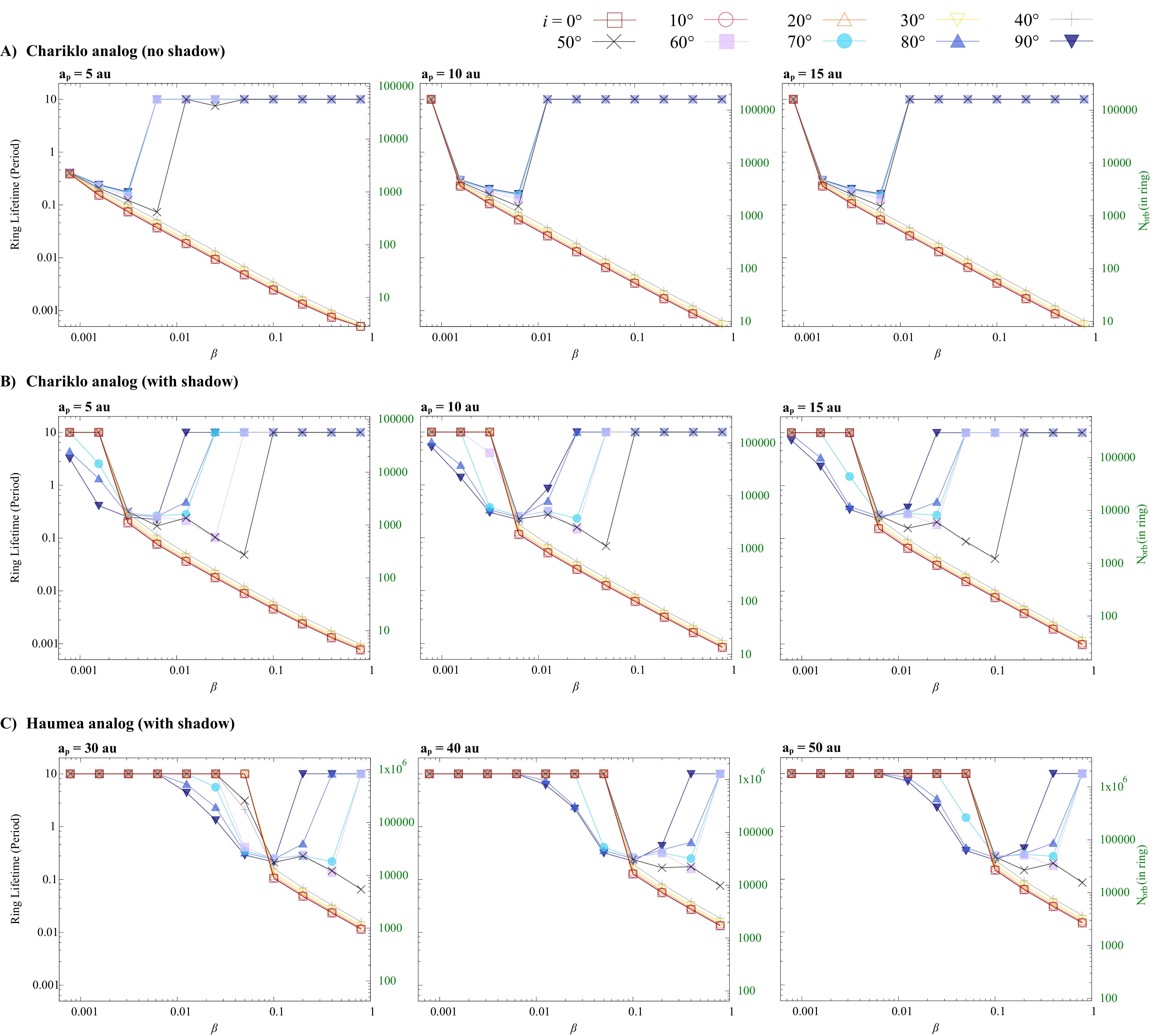}
    \caption{Ring lifetime measured in terms of the orbital period of the planetary body as a function of $\beta$.
    Symbols represent different initial ring tilt angles in the range of $[0^\circ-90^\circ]$.
    The number of orbits completed by the ring particles around the planetary body are also shown on the right vertical axis with green.
    Three different sets of models are considered. 
    Panel~A shows the Chariklo analog models without taking into account the planetary shadow, assuming 5, 10, and 15~au orbital distance.
    Panel~B shows the same models but considering the planetary shadow.
    Panel~C shows the Haumea analog models taking into account the effect of planetary shadow, assuming orbital distances of 30, 40, and 50~au.
    It is noticeable in all panels that $i<50^\circ$ (warm colored symbols) and $i\geq50^\circ$ (cool colored symbols) represent two different groups.
    } 
    \label{fig:ring-lifetime}
\end{figure*}

\begin{figure*}
    \centering
    \includegraphics[width=0.9\linewidth]{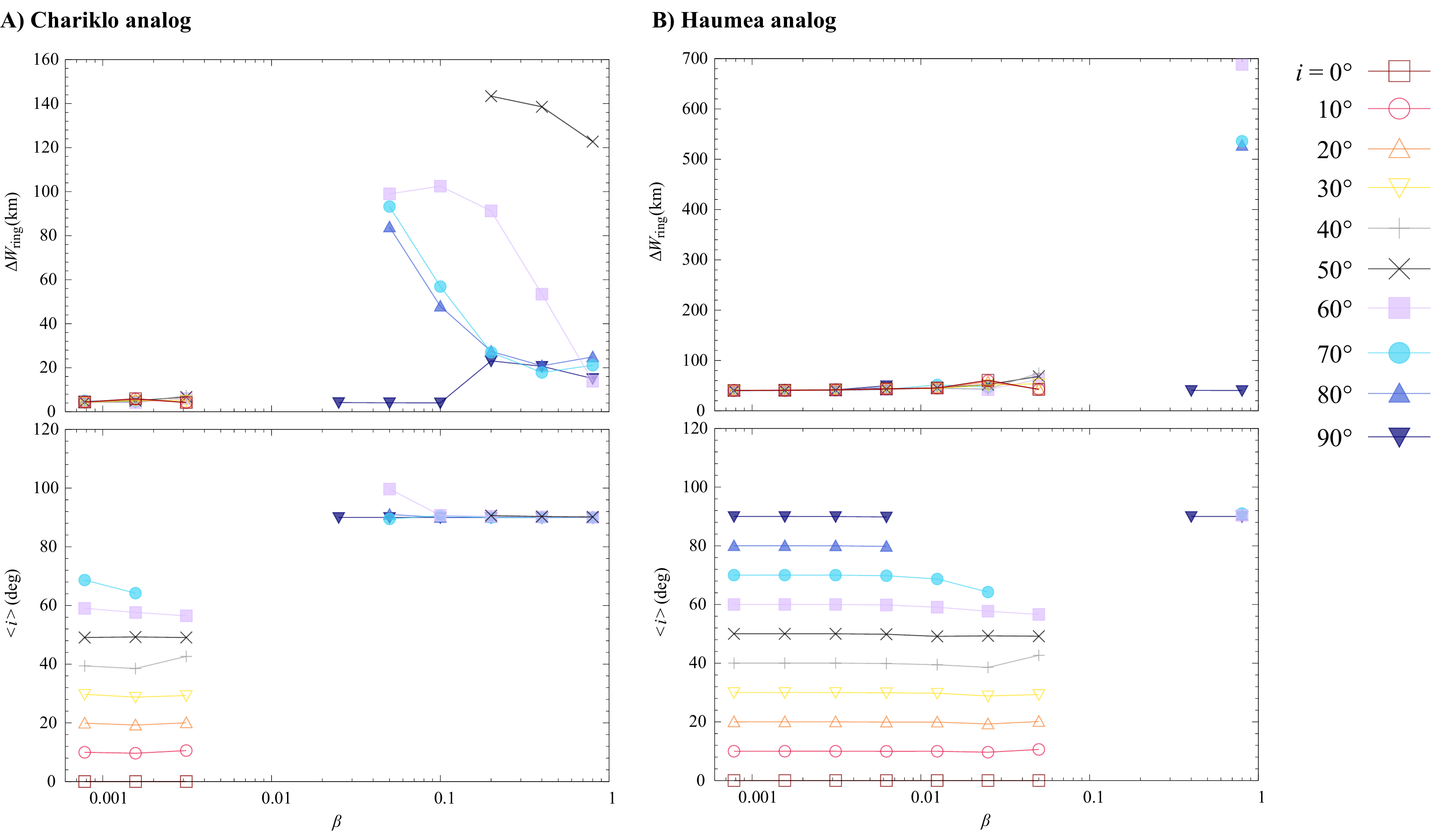}
    \caption{
    Width of rings derived from the valuer of $\Delta W_\mathrm{ring}$ (upper panels) and the average inclination angle of the particles in the ring (lower panels) as a function of $\beta$.
    Panels~A and B show the Chariklo ($a_\mathrm{p}=15$~au) and Haumea ($a_\mathrm{p}=50$~au) analog models, respectively.
    Symbols represent different initial ring tilt angles in the range of $[0^\circ-90^\circ]$.}
    \label{fig:ring-prop}
\end{figure*}

\subsection{Ring lifetime}

The measured ring lifetime is presented as a function of the $\beta$ parameter for a range of models assuming different initial ring tilt angles, orbital distances, and minor body masses in Fig.~\ref{fig:ring-lifetime}.
Panel~A shows models of the Chariklo analog where the planetary shadow is neglected.
All coplanar rings ($i=0^\circ$, shown by dark red symbols) become unstable above a critical $\beta_\mathrm{crit}\gtrsim10^{-3}$ in all models.
This $\beta_\mathrm{crit}$ increases with the orbital distance of the planetary body.
This can be simply explained by the weakening of the radiation pressure as the planetary body orbits farther from the Sun, while the planetary gravitational well remains the same.
As long as the ring tilt angle is $i\leq40^\circ$, ring lifetime shows a similar behavior to the coplanar case with a slightly increased lifetime for larger inclinations.
It is important to note that in this case the rings' lifetime is shorter than one orbit of the planetary body around the Sun.

For highly inclined rings ($i\geq50^\circ$), ring stabilization occurs at relatively high values of $\beta_{\mathrm{crit}}^*\gtrsim10~ \beta_\mathrm{crit}$; this is discussed in more detail in Sect.~\ref{sec:Disc}).
Independent of the ring tilt angle, rings stabilize above the same $\beta_{\mathrm{crit}}^*$
As a result, the lifetime of highly tilted rings can exceed the length of the simulations.
The newly defined $\beta_{\mathrm{crit}}^*$ also increases with orbital distance. 
This means that for highly inclined rings, there exists a region of instability between $\beta_{\mathrm{crit}}$ and $\beta_{\mathrm{crit}}^*$.

Comparing models where the planetary shadow is taken into account for the Chariklo analogs (panel B of Fig.~\ref{fig:ring-lifetime}), it is clearly seen that ring lifetime is strongly affected by the planetary shadow, in contrast to previous assumptions in the literature.
In general, the effect of shadow is to lengthen the ring lifetime as well as to increase $\beta_\mathrm{crit}$ by half an order of magnitude.
Another effect of the shadow is that the instability region for the initially highly tilted ($i\geq50^\circ$) rings widens.
The rings are unstable for $0.1~\beta_\mathrm{crit}\lesssim\beta\lesssim10~\beta_\mathrm{crit}$, and this range shifts toward larger $\beta$ values for larger orbital distances of the planetary body.
We emphasize that the lowest investigated value of $\beta$ is still incapable of stabilizing the ring for $i\geq70^\circ$ due to shadowing.
Generally, these phenomena highlight the importance of the planet's shadow-casting effect when calculating the orbital dynamics of ring particles.

Panel~C of Fig.~\ref{fig:ring-lifetime} shows models of the Haumea analog, where the planetary shadow is taken into account.
As one can see, $\beta_\mathrm{crit}\simeq10^{-1}$ for $i\leq40^\circ$.
Moreover, the stability region of highly inclined rings is also shifted toward higher $\beta$.
The dependence of $\beta_\mathrm{crit}$ on the orbital distance of the planetary body is much weaker than for the Chariklo analog.

As a summary, the numerical results agree well with the theoretical predictions of $\beta_\mathrm{crit}$ as long as the shadow is neglected and the initial ring tilt angle is $i\leq40^\circ$.
However, due to the non-negligible effect of the planetary shadow, $\beta_\mathrm{crit}$ is about five times larger in cases where the shadow is modeled.
For high initial ring tilt angles, $i\geq50^\circ$, the ring can be unstable for a wide range of $\beta$, $0.1~\beta_\mathrm{crit} \lesssim \beta  \lesssim10~\beta_\mathrm{crit}$.
Nevertheless, for large enough $\beta\gtrsim 10~\beta_\mathrm{crit}$, the initially tilted rings can be stabilized.

\subsection{Geometry of surviving rings}

Let's now discuss a method for estimating the width of the surviving rings in order to determine its geometric properties.
The radial width of a planetary ring, $\Delta W_\mathrm{ring}$, is derived by considering the statistical dispersions in the orbital elements of its constituent particles. 
The semimajor axes of the particles, $a$ are distributed around a mean value, $\left< a \right>$, with a standard deviation $\sigma_a$. 
This results in a radial variation of approximately $2\sigma_a$, representing the full-width contribution from the semimajor axis dispersion. 
Additionally, the eccentricity, $e$, of the particles introduces further variation, as the radial excursion of a particle's orbit is proportional to both its semimajor axis and eccentricity as $r = a(1 \pm e)$. 
Considering the dispersion in eccentricity, $\sigma_e$, the corresponding radial contribution is approximated as $2\left< a  \right>\sigma_e$. 
Combining these independent contributions, the total radial width of the ring is expressed as 
\begin{equation}
\Delta W_\mathrm{ring} = 2\sigma_a + 2\left<a\right>\sigma_e.
\end{equation}
This derivation assumes that the dispersions in $a$ and $e$ are uncorrelated and approximately Gaussian, enabling a simple linear combination of their effects to estimate the ring’s radial extent. 

Figure~\ref{fig:ring-prop} shows the geometric properties of stable rings at the end of the simulations for Chariklo and Haumea analogs.
In these models, the semimajor axis of the solar orbit is $a_\mathrm{p}=15$~au and $50$~au, respectively.
For the Chariklo analog (top plot in panel~A), assuming low initial ring tilt angles and $\beta\leq3.125\times10^{-3}$ (corresponding to $\gtrsim5-70~\mu$m depending the composition of the ring particles), the width of the ring does not change and remains about 10~km.
We note that for larger $\beta$ there are no stable low inclination rings.
However, for large tilt angles the ring expands to 100~km for intermediate values, $0.01\leq\beta\leq0.1$ ($0.5~\mu\mathrm{m}\lesssim s \lesssim25~\mu\mathrm{m}$ particles) and about 20~km for large values, $\beta>0.1$ ($s\lesssim2~\mu\mathrm{m}$).
For the Haumea analog (top plot in panel B), the ring width does not change and remains at 70~km for $\beta\leq5\times10^{-2}$ (corresponding to $1-4.5~\mu\mathrm{m}$ depending on the composition of the ring particles), assuming small initial ring tilt angles.

Based on the calculated average tilt of the ring particles (bottom plot in panel~A), it is noticeable that the ring tilt angle does not change significantly for $\beta\leq3.125\times10^{-3}$ (corrsponding to $\gtrsim 15-70~\mu\mathrm{m}$ depending on the composition of the ring particles) in the case of the Chariklo analog.
For this range of $\beta$ there are no stable rings with an initially high tilt angle ($i>70^\circ$).
For Haumea analogs (bottom plot in panel~B), however, the $\beta$ range is extended to about $5\times10^{-2}$.
It is noticeable that the maximum $\beta$ for stable rings decreases with initial ring tilt angles.

In both model sets, for initially highly tilted ($i\geq50^\circ$) rings, there is evidence for an interesting phenomenon responsible for ring stabilization.
In the Chariklo analogous model it is striking that for $\beta\geq2.5\times10^{-2}$ (corresponding to $\lesssim2-10~\mu$m depending on the composition of ring particles) the ring becomes aligned such that $\left<i\right>=90^\circ$ by the end of the simulation.
For the Haumea analog, this alignment is only possible at a very high value of $\beta$, which is greater than 0.4. This can occur only for amorphous carbon composition with a size of approximately larger than $0.7~\mu$m (see Fig.~\ref{fig:beta-size}).
As a result of the alignment of the rings, the orbital motion of the particles is ordered to be perpendicular to the solar radiation.
This phenomenon will be the subject of discussion below.

\begin{figure*}[h!]
    \centering
    \includegraphics[width=1\linewidth]{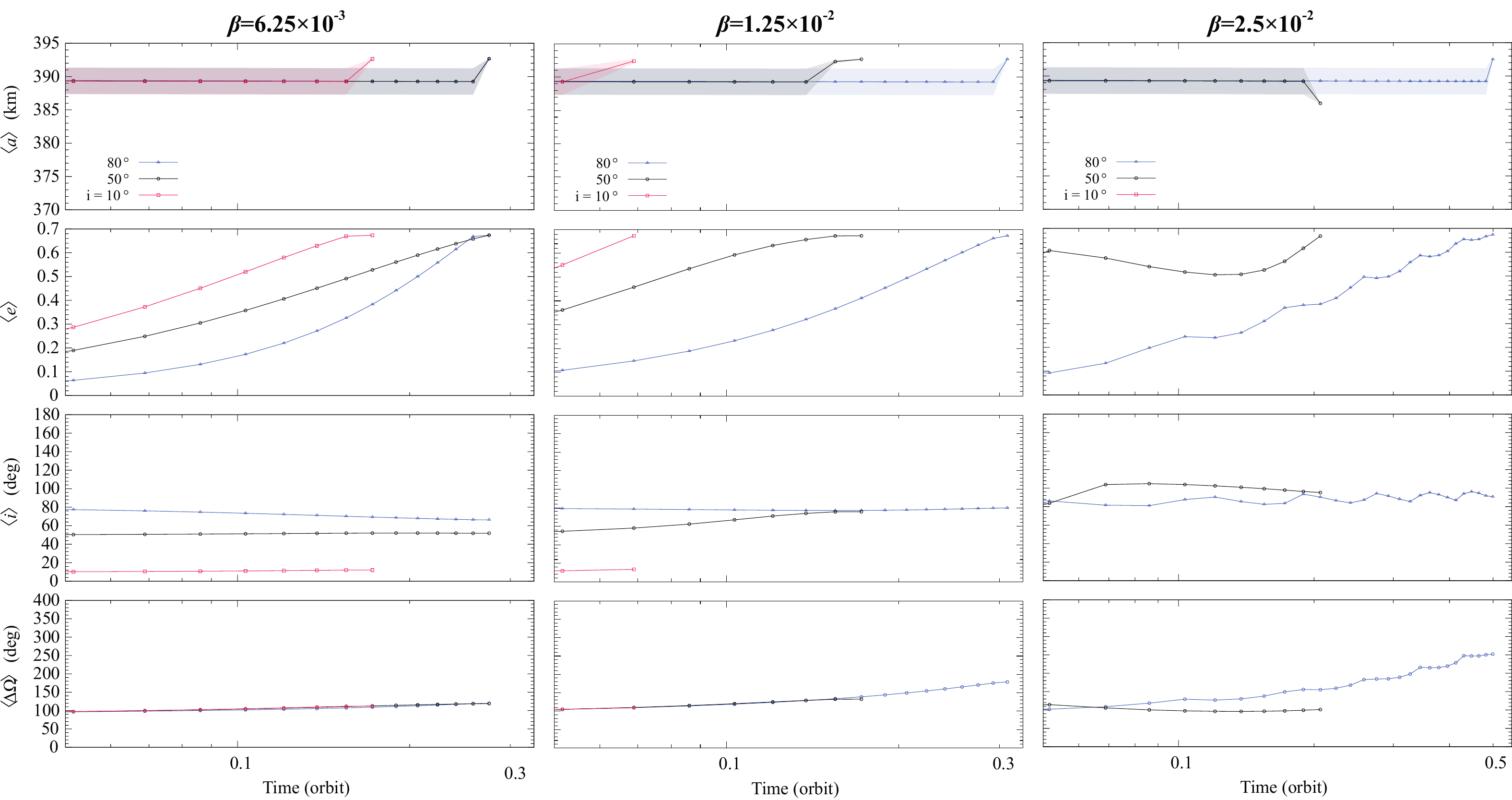}
    \caption{Evolution of average orbital parameters (from top to bottom: $\left<a\right>$, $\left<e\right>$, $\left<i\right>$, and $\left<\Omega\right>$) of ring particles as a function of time, measured in the orbital period of the planetary body assuming $\beta>\beta_\mathrm{crit}$.
    Here we present a Chariklo analog with $a=15$~au.
    Shaded regions represent the $1\times\sigma$ deviation from the average.
    The effect of planetary shadow has been considered in all models presented here.
    }
    \label{fig:orbelems-B3-B5}
\end{figure*}

\section{Discussion}

\subsection{Evolution of orbital elements}
\label{sec:Disc}

\begin{figure*}
    \centering
    \includegraphics[width=1\linewidth]{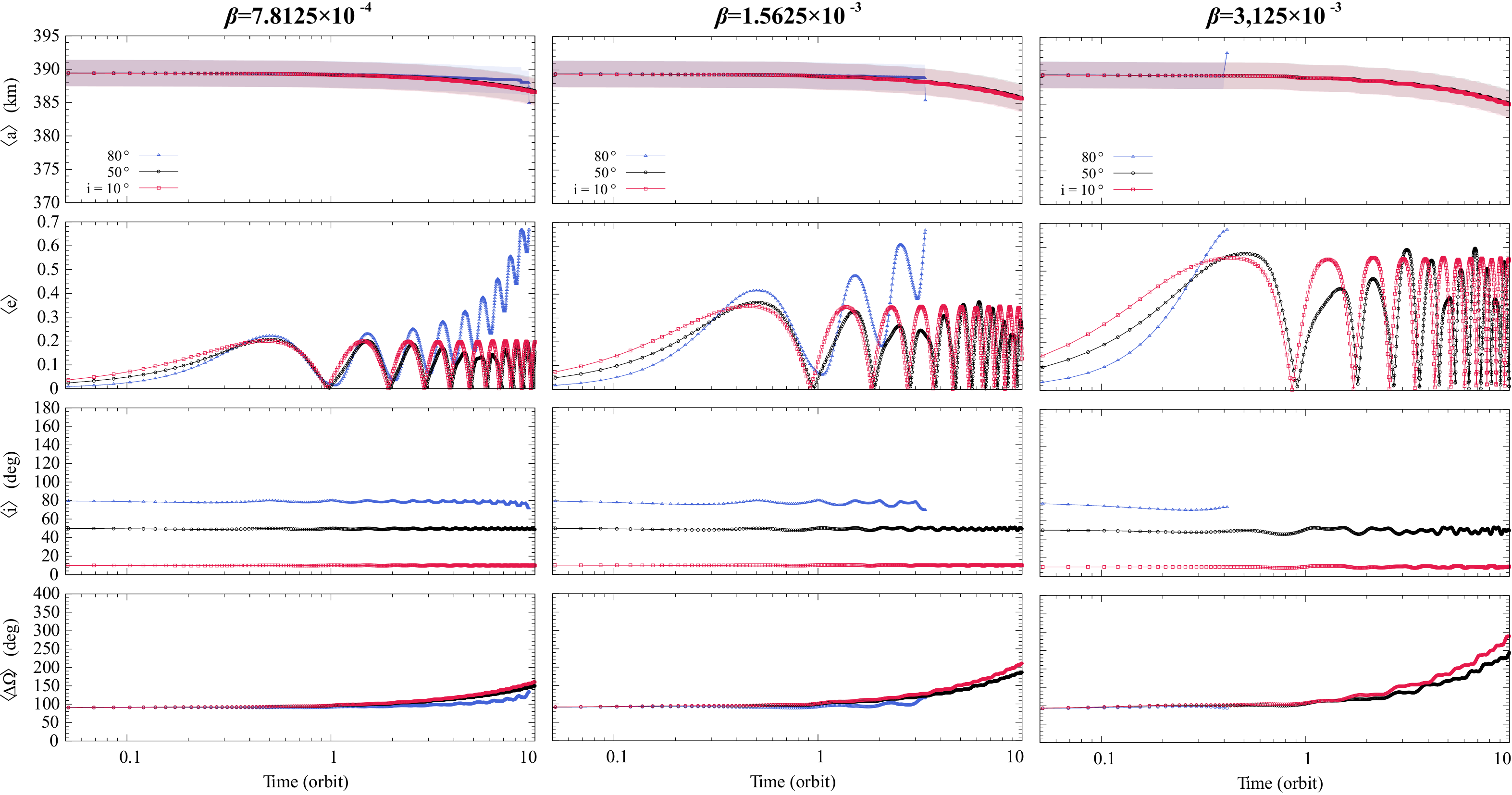}
    \caption{Same as Fig.~\ref{fig:orbelems-B3-B5} but $\beta<\beta_\mathrm{crit}$ is assumed. 
    The ring is unstable for $i=80^\circ$ while it is stable for $i=10^\circ$ and $i=50^\circ$.
    The effect of planetary shadow has been considered in all models presented here.}
    \label{fig:orbelems-B0-B2}
\end{figure*}

\begin{figure*}
    \centering
    \includegraphics[width=1\linewidth]{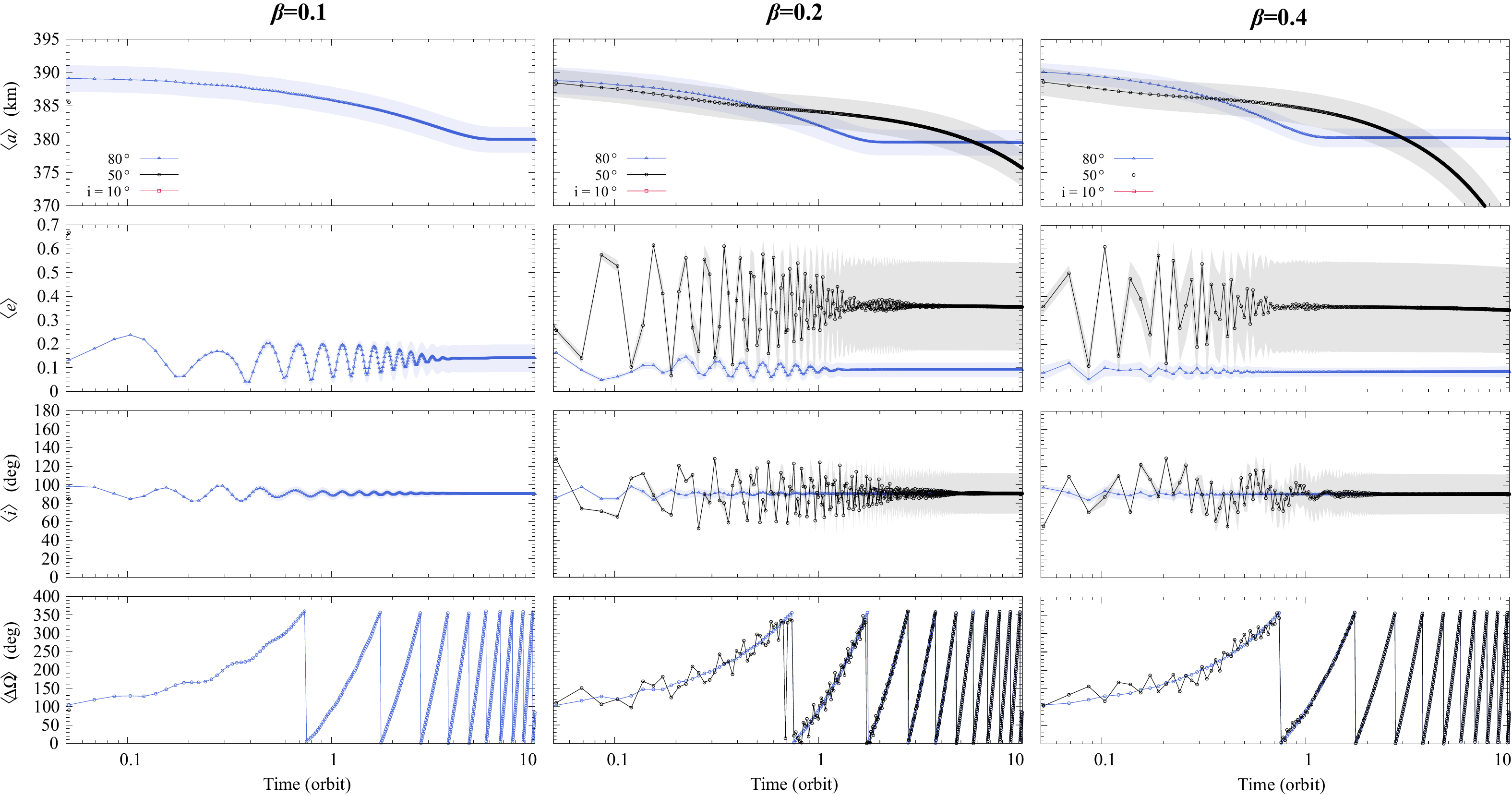}
    \caption{Same as Fig.~\ref{fig:orbelems-B3-B5} but for $\beta>10~\beta_\mathrm{crit}$.
    The $i\geq50^\circ$ models represent stable ring configurations.
    The effect of planetary shadow has been considered in all models presented here.}
    \label{fig:orbelems-B6-B8}
\end{figure*}

To comprehend the influence of the initial ring conditions on ring lifetime, we examine the progression of the averaged orbital elements.
The orbital elements are determined using the relative positions and velocities of the ring particles with respect to the planetary body.
To this end, we calculated the evolution of the particles' average semimajor axis, $\left<a\right>$, eccentricity, $\left<e\right>$ inclination, $\left<i\right>$, and longitude of ascending node, $\left<\Omega\right>$, using regularly generated snapshots of the simulations.
The evolution of the averaged orbital parameters are investigated for three initial ring tilt angles: $i=10^\circ,~50^\circ$, and $80^\circ$. 

First, let's examine models with $\beta>\beta_\mathrm{crit}$, which results in unstable rings regardless of the initial ring tilt angle.
Figure~\ref{fig:orbelems-B3-B5} shows examples of the evolution of the orbital elements for three different $\beta$ parameters assuming a Chariklo analog. 
It is appreciable that the eccentricity quickly reaches the critical value in less than $0.5$ orbit periods of the planetary body in all models.
During this time the average semimajor axis and the inclination angles of the ring particles do not change significantly.
We note that the orbital parameters do not show any measurable scattering across the ring ($\sigma_e,~\sigma_i$ and $\sigma_\Omega$ are negligibly small), meaning that all particles have the same orbital parameters.
An exception is the orbital distance of the ring particles from the planetary body (see upper panels), which is due to the finite (10~km) thickness of the ring.

Figure~\ref{fig:orbelems-B0-B2} shows the examples of the evolution of the averaged orbital parameters of the ring particles with $\beta<\beta_\mathrm{crit}$.
As can be seen, only the highly tilted models ($i=80^{\circ}$) are unstable. 
For smaller initial tilt angles ($i\leq50^\circ$), the ring is stable and the eccentricity of the particles oscillates. 
Although the maximum eccentricity value increases with $\beta$, it stays below the critical value, which explains the stability of the ring.
We note, however, that the ring is slowly approaching the planetary body.
As a result, the ring will be engulfed by the planet over a longer time scale.
Unlike the previous cases, the tilt of the ring also oscillates with time, but with an insignificant amplitude.
On the other hand, the longitude of the ascending node increases slightly: the eccentric ring rotates at a faster rate for larger $\beta$, independent of~$i$.

Figure~\ref{fig:orbelems-B6-B8} shows that in models where the ring consists of particles with $\beta\geq\beta_\mathrm{crit}^*$, the ring is stabilized for sufficiently high initial tilt angles ($i\geq50^\circ$).
As can be seen, at $i\geq50^{\circ}$ the ring is stable.
Interestingly, the ring stabilizes at a distance of 380~km in the most highly inclined model independently of $\beta$. 
Stabilization happens on a shorter timescale for larger $\beta$ values.
Contrary to the previous cases, the strength of the oscillation in the eccentricities and inclination of the ring particles decreases with time and ceases within 10 orbits of the planetary body.
We note that the particles do not possess the same eccentricity and inclination as a relatively large deviation from the average can be observed. 
The scatter in eccentricity and inclination is smaller for larger initial tilt angle of the ring.
We emphasize that the final $\left<i\right>$ values tend to approach $90^\circ$ for all stable cases.

An interesting phenomenon can be observed in the evolution of the longitude of the ascending node: the ring plane rotates.
By comparing the direction of solar radiation and that of $\Omega$, we find that the ring plane is always perpendicular to the direction of the solar radiation.
This behavior of the ring is similar to that of a sunflower\footnote{Young sunflower plants (Helianthus annuus) exhibit heliotropism, meaning their flower buds and leaves turn to face the Sun as it moves across the sky. This behavior maximizes their exposure to sunlight, enhancing photosynthesis. However, once the flower matures, it usually stabilizes facing east and no longer tracks the Sun, but this behavior is a prominent characteristic of the young plants.}, which is why we may imagine highly tilted rings as Celestial Sunflowers.

\subsection{Comparison to known ring systems}

The particle sizes in the rings of small Solar System bodies 
are inferred from observational constraints and theoretical models.
For Chariklo, the estimated particle sizes range from submicron to a few millimeters, consistent with the sharpness and photometric properties of its rings, which are composed of 20\% of water ice, 40-60\% of silicates, and 10-30\% of tholins and small quantities of amorphous carbon \citep{BragaRibas2014,Duffard2014A&A...568A..79D}.
Chiron’s putative ring system, inferred from brightness variations likely consists of water ice \citep{Ortizetal2015A&A...576A..18O}.
Haumea’s ring, observed through stellar occultation has a reflectivity of $0.09$, which is is comparable to the main ring of Chariklo \citep{Braga-Ribasetal2014Natur.508...72B,Leiva2017AJ....154..159L,Ortiz2017}.
\citet{Kalup2024PASP..136l4401K} has recently investigated radiative transfer models for future observations assuming different materials and particle sizes (<1mm) for Haumea's ring.
Quaoar’s ring, recently detected via occultation, suggests a distribution dominated by micron- to millimeter-sized particles, supported by its sharp boundaries and collisional dynamics \citep{Morgadoetal2023Natur.614..239M}.
These grain size distributions are shaped by radiation pressure, collisional processes, and dynamical stability, with future observations poised to refine these estimates.

Thorough analysis of the occultation data of Haumea revealed that its ring plane is inclined by only $3.2^\circ\pm1.4^\circ$ with respect to the planetary body's orbital plane
\citep{Kondratyev2018MNRAS.478.3159K}.
The recently discovered two rings around Quaoar are also in Quaoar's orbital plane.
Thus, these TNOs resemble our low inclination models for which case the ring stability requires $\beta\lesssim5\times10^{-2}$, see panel~C of Fig.~\ref{fig:ring-lifetime}.
This corresponds to a grain size of $s\gtrsim 1-4~\mu$m depending on the composition of the grains, see Fig.~\ref{fig:beta-size}.

The occultation data of Chariklo and Chiron suggest relatively high tilt angles of the rings with respect to their orbital plane.
The occultation observation taken on 29/11/2011 of Chariklo showed that its ring is tilted by $\sim86^\circ$ \citep{Morgado2021A&A...652A.141M}.
The two occultation observations of Chiron on 29/11/2011 and 15/12/2022 showed that its ring tilt is slightly changed from $\sim59^\circ$ to $\sim68^\circ$ \citep{Ortizetal2023A&A...676L..12O}.
Based on our results, there are two possible regions of $\beta$ values for which the ring is stable: 
either for small particles with $\beta\gtrsim0.02$ or for large particles with $\beta\lesssim5\times10^{-4}$, see panel~B of Fig.~\ref{fig:ring-lifetime}.
Depending on the composition of the particles these $\beta$ values correspond to $s\lesssim2.5-15~ \mu$m for small particles or to $s\gtrsim60-300~\mu$m for large particles, see Fig.~\ref{fig:beta-size}.
It has also been suggested that very small -- therefore short-lived -- grains may be responsible for the mid-infrared excess emission observed from the dwarf planet Makemake \citep{Kiss2024}.

Finally, let us say a few words about the ring systems of the giant planets in the Solar System. 
The ring systems of Jupiter, Saturn, and Neptune all lie close to the orbital plane of their respective planets, which means that the $i\leq40^\circ$ models are applicable to them. 
The masses of these giant planets exceed those of the planetary bodies examined here by six to nine orders of magnitude. 
As we have seen in Fig.~\ref{fig:ring-lifetime}, an increasing planetary mass causes the critical $\beta$ to shift toward very large values. 
Therefore, it can be stated that these giant planets' rings may contain smaller than micrometer–sized dust particles.

The ring system of Uranus is nearly perpendicular to its orbital plane, so we must consider the simulations of the $i\geq50^\circ$ ring family. 
Since the mass of Uranus is four to seven orders of magnitude greater than that of the planetary bodies examined here, it can also be stated that the instability region shifts significantly toward higher $\beta$ values.
Additional simulations of the ring of Uranus at a distance of $66,100$~km from the planet with a ring tilt angle of $i=80^\circ,~90^\circ$ revealed that that rings are stable for $\beta\leq0.2$.
Consequently, the small-particle limit of the ring stability range is found to be around 0.1-1 micrometers depending on the composition of ring particles. 
According to these additional simulations, the ring does not show any sunflower-like behavior, thus the ring keeps its plane orientation unchanged as Uranus orbits the Sun.
This is because the gravitational well of Uranus is much deeper than the small bodies investigated here.
We note that the $\mu$, $\nu$, $\lambda$ and 1986U2R/$\zeta$ rings of Uranus are known to be composed of micrometer-sized dust \citep{Ockert1987JGR....9214969O, Smith1986Sci...233...43S, Showalter2006DPS....38.3808S}.

Several observational predictions can be made from the reach dynamics of tilted rings.
First, stable rings consisting of relatively small particles (for which case $\beta$ is high) around small bodies need not be circular, but may be eccentric due to eccentricity excitation.
This phenomenon occurs for relatively small particles where the effect of radiation pressure is significant, see panel B in Fig.~\ref{fig:snapshots}.
In this case, the small body is not in the geometric center of the elliptical ring, rather it is in one of the focal points.
Another prominent effect of the radiation pressure is that the ring can become very thick for highly tilted cases, see panel C in Fig.~\ref{fig:snapshots}.
As a result, a highly tilted ring can produce peculiar observational effects such as wide occultations.
Moreover, in special viewing configurations where the ring is observed edge-on (see panel C or B in Fig.~\ref{fig:snapshots}), only a single wide occultation event occurs.

\subsection{Caveats}
\label{sec:caveats}

There are a few caveats of the model presented here.
First, we neglect the Pointing-Robertson (PR) drag.
The PR drag force acts in the opposite direction
to the particle velocity, which removes energy and angular momentum from the particle, reducing its semimajor axis and eccentricity.
According to \citet{Burnsetal1979Icar...40....1B}, the orbital decay timescale of particles orbiting a planet is $\simeq530 a/\beta$ years, assuming that the orbital planes of the ring and the planet coincide.
This corresponds to an orbital collapse on a timescale of a hundred thousand years for $\beta=1$.
Since we have modeled 10 orbits of the planetary body, which corresponds to a maximum of several thousand years, our results are not affected by the PR drag.
We note, however, that the study of the long-term stability of rings requires the inclusion of PR drag.

We have also neglected the effect of the Lorentz force acting on charged grains in the magnetic field.
This seems a plausible simplification, since small planetary bodies hardly have significant magnetic fields.
Plasma drag, which could increase the orbital energy and the semimajor axis of the ring particles, has also been neglected.
Since in our case the distance of the ring from the planetary body is only a few diameters, these simplifications seem plausible, see for example Fig~2 of \cite{Sachse2018Icar..303..166S}. 

We assume that all bodies in the system have spherical symmetry.
It is a valid assumption for the Sun (due to its distance) and the particles (due to their size with respect to the ring dimensions).
However, in reality, small planetary bodies are usually not spherically symmetric.
rather, their shape is a triaxial ellipsoid.
Thus, the asymmetric gravitational potential of the planetary body can cause additional perturbations to the ring particles' orbit \citep{Sicardy2020AJ....159..102S}.
The effect of a spherical body with mass anomaly or nonspherical potential on ring dynamics have been investigated in detail (see, e.g., \citealp{Sachse2018Icar..303..166S, Sicardy2019NatAs...3..146S,Madeira2022MNRAS.510.1450M,Ribeiro2023MNRAS.525...44R}).
We are enhancing the capabilities of our N-body code to model arbitrarily shaped small bodies. 
Our preliminary results show that slightly tilted rings can be stabilized by the triaxial shape of the small body.
However, the biaxial shape of the gravitational harmonics of a triaxial body (studied in \citealp{Sicardy2019NatAs...3..146S}) can disrupt the sunflower mechanism.
Whether the ring stabilization mechanism is still operational around nonspherical small bodies will be investigated in a forthcoming study.

Chariklo and Haumea change their distance from the Sun periodically because they have eccentric orbits with $e=0.1721$ and $e=0.19642$, respectively.
In this study, however, we have examined planetary bodies in circular orbits.
We showed that the orbital distance of the planetary body has only a minor effect on the ring dynamics via changing the critical $\beta$ parameters changes slightly.
Nevertheless, for a precise prediction of the dynamics of planetary bodies' ring systems, it will be necessary to take the orbital eccentricities of the planets into account.
Moreover, the strong perturbations from giant planets during close encounters was also neglected.

We neglect the collisions of the ring particles, which could decrease the overall size of the ring particles on the collisional timescale.
As a result, the particle size in the ring eventually reaches the critical $\beta$.
Since coplanar rings are unstable above the critical $\beta$, it is expected that ring stability on a million-year long timescale requires a highly tilted ring such that its plane is perpendicular to the radiation pressure.

Finally, we do not consider the possibility of moons orbiting the planetary bodies.
It is known that Haumea and Quaoar possess two and one small moons, respectively.
Recently, \citet{Sickafoose2024PSJ.....5...32S} have shown that the ring around Chariklo can be stabilized by a small prograde moon if it is in orbital resonance with the ring and the mass of the moon is at least a few times $10^{13}$~kg.
We note, however, that the gravitational effect of shepherding moons is relatively weak compared to the radiation pressure.
Namely, the shepherding moon causes secular perturbation while the perturbation from radiation pressure is fast. 
As a result, slightly inclined rings consisting of small particles presumably cannot be stabilized by shepherding moons.
Conversely, for highly inclined rings, a shepherding moon can have a serious effect on the sunflower–like dynamics.
This is because the tilt of the ring increases to $\sim90^\circ$ over time, while the moon's orbital plane stays the same.
Thus, incorporating a shepherding moon with a prograde or retrograde orbit may be necessary in a future study.

\section{Conclusion}

In this paper, we studied the effect of solar radiation pressure on the stability of a ring around planetary bodies.
Based on the work of \citet{Burnsetal1979Icar...40....1B}, we presented an analytical prediction of the critical particle size for which rings around planetary bodies are stable.
This analytical model assumes that the ring is in the planetary body's orbital plane and neglects the effects of the planetary shadow cast on the ring.
To study the dynamics of tilted rings and the effect of the planetary shadow, we performed N-body simulations using a high-precision 8th-order Hermite GPU-based integrator with 132,000 particles.
We explored a wide range of particle sizes by assuming a ratio of solar gravitation to radiation force of $7.8125\times10^{-4}\leq\beta\leq0.8$ and an initial ring tilt angle (with respect to the orbital plane of the planetary body) in the range of $0^\circ$ to $90^\circ$.
We studied Chariklo and Haumea analogs which are different by nearly three orders of magnitude in mass and are assumed to be spherically symmetric.
The planetary body analogs were assumed to be in a circular orbit around the Sun with different semimajor axes.
The numerical simulations span several hundred and thousand years for Chariklo and Haumea analogs, respectively.
As a result of the solar radiation pressure, the ring particles gain eccentricity depending on their size.
Above a critical particle size, the eccentricity of the particles can grow large enough to destabilize the ring by being accreted by the planetary body.
The stability of the ring on a millennial timescale is characterized by measuring the time it takes to lose 99.9 percent of the particles.
It is considered stable if no particle loss is measured until the end of the simulation, which lasts ten orbits of the planetary body around the Sun.
Our main findings are the following.

\begin{enumerate}
    \item 
    In shadowless simulations, two families of rings have been found depending on the initial tilt angle of the ring.
    The critical angle separating the two families of different dynamical characteristics is found to be between $40^\circ-50^\circ$.
    In the first family, the slightly tilted rings with  $i\leq40^\circ$, are unstable when made up of small particles and are stabilized for large particles.
    In the second family, highly tilted rings with $i\geq50^\circ$ are unstable for a range of particle sizes, which spans 1-10 times the critical size.
    Outside of this range the highly tilted rings are stable.

    \item
    The shadow of a planetary body has a significant effect on the dynamics of the ring particles, even though the two families can still be clearly identified. 
    The critical size of particles that can survive in the slightly tilted ($i\leq40^\circ$) rings is reduced by a factor of about five due to the planet’s shadow casting compared to the analytic prediction of \citep{Burnsetal1979Icar...40....1B}.
    The instability region for highly tilted rings ($i\geq50^\circ$) extends to about 0.1-10 times the newly identified critical particle size due to the effect of the planetary shadow.

    \item 
    The critical particle size decreases with the orbital distance and the mass of the planetary body assuming the same initial ring radius.
    In the first family, the lifetime of the rings does not depend significantly on the tilt angle of the rings.

    \item 
    Stabilization in the second family occurs because of radiation pressure, which renders the tilt angle of the ring to be $90^\circ$.
    Moreover, the plane of the ring is continuously rotated in time so that it is always perpendicular to the direction of solar radiation, analogous to the behavior of sunflowers.
\end{enumerate}

As a final thought, particle size in slightly tilted rings observed around spherically symmetric Chariklo and Haumea analogs should be, depending on composition, above 15-70~$\mu$m and 1-4~$\mu$m, respectively.
For highly tilted rings, there are two plausible particle size ranges that can be stable for over a millennium (corresponding to ten orbits of the planetary bodies).
Rings that contain particles with sizes below 2.5-15~$\mu$m for Chariklo and below 0.08-0.3~$\mu$m for Haumea show sunflower-like behavior. 
However, rings can also be stable if their overall particle size is above 60-300~$\mu$m for Chariklo and above 9-40~$\mu$m for Haumea analogs depending on composition.

\begin{acknowledgements}
VF is supported by the undergraduate research assistant program of the Konkoly Observatory.
ZR acknowledges the helpful discussion on the ring particle properties of solar system small bodies with Cs. Kalup.
We are grateful to the anonymous reviewer for the improvement of our manuscript.
\end{acknowledgements}

\bibliographystyle{aa}
\bibliography{CelestialSunflowers}

\begin{thebibliography}{76}
\expandafter\ifx\csname natexlab\endcsname\relax\def\natexlab#1{#1}\fi

\bibitem[{Allan(1967)}]{Allan1967}
Allan, R. 1967, Planetary and Space Science, 15, 53

\bibitem[{Allan(1962)}]{Allan1962SatelliteOP}
Allan, R.~R. 1962, Quarterly Journal of Mechanics and Applied Mathematics, 15,
  283

\bibitem[{Bohren \& Huffman(1983)}]{bohren1983}
Bohren, C.~F. \& Huffman, D.~R. 1983, Absorption and Scattering of Light by
  Small Particles (Wiley)

\bibitem[{{Braga-Ribas} {et~al.}(2014){Braga-Ribas}, {Sicardy}, {Ortiz},
  {Snodgrass}, {Roques}, {Vieira-Martins}, {Camargo}, {Assafin}, {Duffard},
  {Jehin}, {Pollock}, {Leiva}, {Emilio}, {Machado}, {Colazo}, {Lellouch},
  {Skottfelt}, {Gillon}, {Ligier}, {Maquet}, {Benedetti-Rossi}, {Gomes},
  {Kervella}, {Monteiro}, {Sfair}, {El Moutamid}, {Tancredi}, {Spagnotto},
  {Maury}, {Morales}, {Gil-Hutton}, {Roland}, {Ceretta}, {Gu}, {Wang},
  {Harps{\o}e}, {Rabus}, {Manfroid}, {Opitom}, {Vanzi}, {Mehret}, {Lorenzini},
  {Schneiter}, {Melia}, {Lecacheux}, {Colas}, {Vachier}, {Widemann},
  {Almenares}, {Sandness}, {Char}, {Perez}, {Lemos}, {Martinez},
  {J{\o}rgensen}, {Dominik}, {Roig}, {Reichart}, {Lacluyze}, {Haislip},
  {Ivarsen}, {Moore}, {Frank}, \&
  {Lambas}}]{Braga-Ribasetal2014Natur.508...72B}
{Braga-Ribas}, F., {Sicardy}, B., {Ortiz}, J.~L., {et~al.} 2014, \nat, 508, 72

\bibitem[{Braga-Ribas {et~al.}(2014)Braga-Ribas, Sicardy, Ortiz,
  {et~al.}}]{BragaRibas2014}
Braga-Ribas, F., Sicardy, B., Ortiz, J.~L., {et~al.} 2014, Nature, 508, 72

\bibitem[{Burns {et~al.}(1979)Burns, Lamy, \& Soter}]{Burns1979}
Burns, J.~A., Lamy, P.~L., \& Soter, S. 1979, Icarus, 40, 1

\bibitem[{{Burns} {et~al.}(1979){Burns}, {Lamy}, \&
  {Soter}}]{Burnsetal1979Icar...40....1B}
{Burns}, J.~A., {Lamy}, P.~L., \& {Soter}, S. 1979, \icarus, 40, 1

\bibitem[{{de Pater} {et~al.}(2006{\natexlab{a}}){de Pater}, {Gibbard}, \&
  {Hammel}}]{dePater2006Icar..180..186D}
{de Pater}, I., {Gibbard}, S.~G., \& {Hammel}, H.~B. 2006{\natexlab{a}},
  \icarus, 180, 186

\bibitem[{{de Pater} {et~al.}(2006{\natexlab{b}}){de Pater}, {Hammel},
  {Gibbard}, \& {Showalter}}]{dePater2006Sci...312...92D}
{de Pater}, I., {Hammel}, H.~B., {Gibbard}, S.~G., \& {Showalter}, M.~R.
  2006{\natexlab{b}}, Science, 312, 92

\bibitem[{{Dencs} \& {Reg{\'a}ly}(2019)}]{DencsRegaly2019MNRAS.487.2191D}
{Dencs}, Z. \& {Reg{\'a}ly}, Z. 2019, \mnras, 487, 2191

\bibitem[{{Dencs} \& {Reg{\'a}ly}(2021)}]{DencsRegaly2021A&A...645A..65D}
{Dencs}, Z. \& {Reg{\'a}ly}, Z. 2021, \aap, 645, A65

\bibitem[{{Dominik} {et~al.}(2021){Dominik}, {Min}, \&
  {Tazaki}}]{Dominik2021ascl.soft04010D}
{Dominik}, C., {Min}, M., \& {Tazaki}, R. 2021, {OpTool: Command-line driven
  tool for creating complex dust opacities}, Astrophysics Source Code Library,
  record ascl:2104.010

\bibitem[{{Draine}(2003)}]{Draine2003ApJ...598.1017D}
{Draine}, B.~T. 2003, \apj, 598, 1017

\bibitem[{{Duffard} {et~al.}(2014){Duffard}, {Pinilla-Alonso}, {Ortiz},
  {Alvarez-Candal}, {Sicardy}, {Santos-Sanz}, {Morales}, {Colazo},
  {Fern{\'a}ndez-Valenzuela}, \& {Braga-Ribas}}]{Duffard2014A&A...568A..79D}
{Duffard}, R., {Pinilla-Alonso}, N., {Ortiz}, J.~L., {et~al.} 2014, \aap, 568,
  A79

\bibitem[{{Dunham} {et~al.}(2019){Dunham}, {Desch}, \&
  {Probst}}]{Dunham2019ApJ...877...41D}
{Dunham}, E.~T., {Desch}, S.~J., \& {Probst}, L. 2019, \apj, 877, 41

\bibitem[{{Ferrari} \& {Brahic}(1994)}]{Ferrari1994Icar..111..193F}
{Ferrari}, C. \& {Brahic}, A. 1994, \icarus, 111, 193

\bibitem[{{French} \& {Nicholson}(2000)}]{French2000Icar..145..502F}
{French}, R.~G. \& {Nicholson}, P.~D. 2000, \icarus, 145, 502

\bibitem[{{Giuliatti Winter} {et~al.}(2023){Giuliatti Winter}, {Madeira},
  {Ribeiro}, {Winter}, {Barbosa}, \&
  {Borderes-Motta}}]{Winter2023A&A...679A..62G}
{Giuliatti Winter}, S.~M., {Madeira}, G., {Ribeiro}, T., {et~al.} 2023, \aap,
  679, A62

\bibitem[{{Hamilton} \& {Krivov}(1996)}]{HamiltonKrivov1996Icar..123..503H}
{Hamilton}, D.~P. \& {Krivov}, A.~V. 1996, \icarus, 123, 503

\bibitem[{{Hasegawa} {et~al.}(1977){Hasegawa}, {Fujiwara}, {Koike}, \&
  {Mukai}}]{Hasegawa1977MmKyo..35..131H}
{Hasegawa}, H., {Fujiwara}, A., {Koike}, C., \& {Mukai}, T. 1977, Memoirs
  Faculty of Sciences University of Kyoto, 35, 131

\bibitem[{{Henning} \& {Stognienko}(1996)}]{Henning1996A&A...311..291H}
{Henning}, T. \& {Stognienko}, R. 1996, \aap, 311, 291

\bibitem[{{Herrick}(1948)}]{Herrick1948PASP...60..321H}
{Herrick}, S. 1948, \pasp, 60, 321

\bibitem[{{Irvine} \& {Baragar}(1971)}]{Irvine1971CaJES...8..523I}
{Irvine}, T.~N. \& {Baragar}, W.~R.~A. 1971, Canadian Journal of Earth
  Sciences, 8, 523

\bibitem[{{Kalup} {et~al.}(2024){Kalup}, {Moln{\'a}r}, \&
  {Kiss}}]{Kalup2024PASP..136l4401K}
{Kalup}, C., {Moln{\'a}r}, L., \& {Kiss}, C. 2024, \pasp, 136, 124401

\bibitem[{{Kane} \& {Li}(2022)}]{Kane2022LPICo2687.3041K}
{Kane}, S.~R. \& {Li}, Z. 2022, in LPI Contributions, Vol. 2687, Exoplanets in
  Our Backyard 2, 3041

\bibitem[{Kimura {et~al.}(1997)Kimura, Kolokolova, \& Mann}]{Kimura1997}
Kimura, H., Kolokolova, L., \& Mann, I. 1997, Astronomy and Astrophysics, 326,
  263

\bibitem[{{Kiss} {et~al.}(2024){Kiss}, {M{\"u}ller}, {Farkas-Tak{\'a}cs},
  {Mo{\'o}r}, {Protopapa}, {Parker}, {Santos-Sanz}, {Ortiz}, {Holler}, {Wong},
  {Stansberry}, {Fern{\'a}ndez-Valenzuela}, {Glein}, {Lellouch}, {Vilenius},
  {Kalup}, {Reg{\'a}ly}, {Szak{\'a}ts}, {Marton}, {P{\'a}l}, \&
  {Szab{\'o}}}]{Kiss2024}
{Kiss}, C., {M{\"u}ller}, T.~G., {Farkas-Tak{\'a}cs}, A., {et~al.} 2024, \apjl,
  976, L9

\bibitem[{{Kitamura} {et~al.}(2007){Kitamura}, {Pilon}, \&
  {Jonasz}}]{Kitamura2007ApOpt..46.8118K}
{Kitamura}, R., {Pilon}, L., \& {Jonasz}, M. 2007, \ao, 46, 8118

\bibitem[{{Kondratyev}(2016)}]{Kondratyev2016Ap&SS.361..389K}
{Kondratyev}, B.~P. 2016, \apss, 361, 389

\bibitem[{{Kondratyev} \& {Kornoukhov}(2018)}]{Kondratyev2018MNRAS.478.3159K}
{Kondratyev}, B.~P. \& {Kornoukhov}, V.~S. 2018, \mnras, 478, 3159

\bibitem[{{Kov{\'a}cs} \& {Reg{\'a}ly}(2018)}]{KovacsRegaly2018MNRAS.479.4560K}
{Kov{\'a}cs}, T. \& {Reg{\'a}ly}, Z. 2018, \mnras, 479, 4560

\bibitem[{{Krivov} {et~al.}(2002){Krivov}, {Kr{\"u}ger}, {Gr{\"u}n},
  {Thiessenhusen}, \& {Hamilton}}]{Krivov2002JGRE..107.5002K}
{Krivov}, A.~V., {Kr{\"u}ger}, H., {Gr{\"u}n}, E., {Thiessenhusen}, K.-U., \&
  {Hamilton}, D.~P. 2002, Journal of Geophysical Research (Planets), 107, 5002

\bibitem[{{Leiva} {et~al.}(2017){Leiva}, {Sicardy}, {Camargo}, {Ortiz},
  {Desmars}, {B{\'e}rard}, {Lellouch}, {Meza}, {Kervella}, {Snodgrass},
  {Duffard}, {Morales}, {Gomes-J{\'u}nior}, {Benedetti-Rossi},
  {Vieira-Martins}, {Braga-Ribas}, {Assafin}, {Morgado}, {Colas}, {De Witt},
  {Sickafoose}, {Breytenbach}, {Dauvergne}, {Schoenau}, {Maquet}, {Bath},
  {Bode}, {Cool}, {Lade}, {Kerr}, \& {Herald}}]{Leiva2017AJ....154..159L}
{Leiva}, R., {Sicardy}, B., {Camargo}, J.~I.~B., {et~al.} 2017, \aj, 154, 159

\bibitem[{Lissauer(1993)}]{lissauer1993planetary}
Lissauer, J.~J. 1993, Annual Review of Astronomy and Astrophysics, 31, 129

\bibitem[{{Madeira} {et~al.}(2022){Madeira}, {Giuliatti Winter}, {Ribeiro}, \&
  {Winter}}]{Madeira2022MNRAS.510.1450M}
{Madeira}, G., {Giuliatti Winter}, S.~M., {Ribeiro}, T., \& {Winter}, O.~C.
  2022, \mnras, 510, 1450

\bibitem[{{Mignard}(1982)}]{Mignard1982Icar...49..347M}
{Mignard}, F. 1982, \icarus, 49, 347

\bibitem[{{Morgado} {et~al.}(2021){Morgado}, {Sicardy}, {Braga-Ribas},
  {Desmars}, {Gomes-J{\'u}nior}, {B{\'e}rard}, {Leiva}, {Ortiz},
  {Vieira-Martins}, {Benedetti-Rossi}, {Santos-Sanz}, {Camargo}, {Duffard},
  {Rommel}, {Assafin}, {Boufleur}, {Colas}, {Kretlow}, {Beisker}, {Sfair},
  {Snodgrass}, {Morales}, {Fern{\'a}ndez-Valenzuela}, {Amaral}, {Amarante},
  {Artola}, {Backes}, {Bath}, {Bouley}, {Buie}, {Cacella}, {Colazo}, {Colque},
  {Dauvergne}, {Dominik}, {Emilio}, {Erickson}, {Evans}, {Fabrega-Polleri},
  {Garcia-Lambas}, {Giacchini}, {Hanna}, {Herald}, {Hesler}, {Hinse},
  {Jacques}, {Jehin}, {J{\o}rgensen}, {Kerr}, {Kouprianov}, {Levine}, {Linder},
  {Maley}, {Machado}, {Maquet}, {Maury}, {Melia}, {Meza}, {Mondon}, {Moura},
  {Newman}, {Payet}, {Pereira}, {Pollock}, {Poltronieri}, {Quispe-Huaynasi},
  {Reichart}, {de Santana}, {Schneiter}, {Sieyra}, {Skottfelt}, {Soulier},
  {Starck}, {Thierry}, {Torres}, {Trabuco}, {Unda-Sanzana}, {Yamashita},
  {Winter}, {Zapata}, \& {Zuluaga}}]{Morgado2021A&A...652A.141M}
{Morgado}, B.~E., {Sicardy}, B., {Braga-Ribas}, F., {et~al.} 2021, \aap, 652,
  A141

\bibitem[{{Morgado} {et~al.}(2023){Morgado}, {Sicardy}, {Braga-Ribas}, {Ortiz},
  {Salo}, {Vachier}, {Desmars}, {Pereira}, {Santos-Sanz}, {Sfair}, {de
  Santana}, {Assafin}, {Vieira-Martins}, {Gomes-J{\'u}nior}, {Margoti},
  {Dhillon}, {Fern{\'a}ndez-Valenzuela}, {Broughton}, {Bradshaw}, {Langersek},
  {Benedetti-Rossi}, {Souami}, {Holler}, {Kretlow}, {Boufleur}, {Camargo},
  {Duffard}, {Beisker}, {Morales}, {Lecacheux}, {Rommel}, {Herald}, {Benz},
  {Jehin}, {Jankowsky}, {Marsh}, {Littlefair}, {Bruno}, {Pagano}, {Brandeker},
  {Collier-Cameron}, {Flor{\'e}n}, {Hara}, {Olofsson}, {Wilson}, {Benkhaldoun},
  {Busuttil}, {Burdanov}, {Ferrais}, {Gault}, {Gillon}, {Hanna}, {Kerr},
  {Kolb}, {Nosworthy}, {Sebastian}, {Snodgrass}, {Teng}, \& {de
  Wit}}]{Morgadoetal2023Natur.614..239M}
{Morgado}, B.~E., {Sicardy}, B., {Braga-Ribas}, F., {et~al.} 2023, \nat, 614,
  239

\bibitem[{Mukai(1989)}]{Mukai1989}
Mukai, T. 1989, Astronomy and Astrophysics, 213, 387

\bibitem[{{Nitadori} \& {Makino}(2008)}]{NitadoriMakino2008NewA...13..498N}
{Nitadori}, K. \& {Makino}, J. 2008, \na, 13, 498

\bibitem[{{Ockert} {et~al.}(1987){Ockert}, {Cuzzi}, {Porco}, \&
  {Johnson}}]{Ockert1987JGR....9214969O}
{Ockert}, M.~E., {Cuzzi}, J.~N., {Porco}, C.~C., \& {Johnson}, T.~V. 1987,
  \jgr, 92, 14969

\bibitem[{{Ortiz} {et~al.}(2015){Ortiz}, {Duffard}, {Pinilla-Alonso},
  {Alvarez-Candal}, {Santos-Sanz}, {Morales}, {Fern{\'a}ndez-Valenzuela},
  {Licandro}, {Campo Bagatin}, \& {Thirouin}}]{Ortizetal2015A&A...576A..18O}
{Ortiz}, J.~L., {Duffard}, R., {Pinilla-Alonso}, N., {et~al.} 2015, \aap, 576,
  A18

\bibitem[{Ortiz {et~al.}(2017)Ortiz, Duffard, Pinilla-Alonso,
  {et~al.}}]{Ortiz2017}
Ortiz, J.~L., Duffard, R., Pinilla-Alonso, N., {et~al.} 2017, Nature, 550, 219

\bibitem[{{Ortiz} {et~al.}(2023){Ortiz}, {Pereira}, {Sicardy}, {Braga-Ribas},
  {Takey}, {Fouad}, {Shaker}, {Kaspi}, {Brosch}, {Kretlow}, {Leiva}, {Desmars},
  {Morgado}, {Morales}, {Vara-Lubiano}, {Santos-Sanz},
  {Fern{\'a}ndez-Valenzuela}, {Souami}, {Duffard}, {Rommel}, {Kilic}, {Erece},
  {Koseoglu}, {Ege}, {Morales}, {Alvarez-Candal}, {Rizos},
  {G{\'o}mez-Lim{\'o}n}, {Assafin}, {Vieira-Martins}, {Gomes-J{\'u}nior},
  {Camargo}, \& {Lecacheux}}]{Ortizetal2023A&A...676L..12O}
{Ortiz}, J.~L., {Pereira}, C.~L., {Sicardy}, B., {et~al.} 2023, \aap, 676, L12

\bibitem[{{Ortiz} {et~al.}(2017){Ortiz}, {Santos-Sanz}, {Sicardy},
  {Benedetti-Rossi}, {B{\'e}rard}, {Morales}, {Duffard}, {Braga-Ribas}, {Hopp},
  {Ries}, {Nascimbeni}, {Marzari}, {Granata}, {P{\'a}l}, {Kiss}, {Pribulla},
  {Kom{\v{z}}{\'\i}k}, {Hornoch}, {Pravec}, {Bacci}, {Maestripieri}, {Nerli},
  {Mazzei}, {Bachini}, {Martinelli}, {Succi}, {Ciabattari}, {Mikuz},
  {Carbognani}, {Gaehrken}, {Mottola}, {Hellmich}, {Rommel},
  {Fern{\'a}ndez-Valenzuela}, {Campo Bagatin}, {Cikota}, {Cikota}, {Lecacheux},
  {Vieira-Martins}, {Camargo}, {Assafin}, {Colas}, {Behrend}, {Desmars},
  {Meza}, {Alvarez-Candal}, {Beisker}, {Gomes-Junior}, {Morgado}, {Roques},
  {Vachier}, {Berthier}, {Mueller}, {Madiedo}, {Unsalan}, {Sonbas}, {Karaman},
  {Erece}, {Koseoglu}, {Ozisik}, {Kalkan}, {Guney}, {Niaei}, {Satir},
  {Yesilyaprak}, {Puskullu}, {Kabas}, {Demircan}, {Alikakos}, {Charmandaris},
  {Leto}, {Ohlert}, {Christille}, {Szak{\'a}ts}, {Tak{\'a}csn{\'e} Farkas},
  {Varga-Vereb{\'e}lyi}, {Marton}, {Marciniak}, {Bartczak}, {Santana-Ros},
  {Butkiewicz-B{\k{a}}k}, {Dudzi{\'n}ski}, {Al{\'\i}-Lagoa}, {Gazeas},
  {Tzouganatos}, {Paschalis}, {Tsamis}, {S{\'a}nchez-Lavega},
  {P{\'e}rez-Hoyos}, {Hueso}, {Guirado}, {Peris}, \&
  {Iglesias-Marzoa}}]{Ortizetal2017Natur.550..219O}
{Ortiz}, J.~L., {Santos-Sanz}, P., {Sicardy}, B., {et~al.} 2017, \nat, 550, 219

\bibitem[{{Palik}(1991)}]{Palik1991}
{Palik}, E.~D. 1991, {Handbook of optical constants of solids II}

\bibitem[{{Pereira} {et~al.}(2023){Pereira}, {Sicardy}, {Morgado},
  {Braga-Ribas}, {Fern{\'a}ndez-Valenzuela}, {Souami}, {Holler}, {Boufleur},
  {Margoti}, {Assafin}, {Ortiz}, {Santos-Sanz}, {Epinat}, {Kervella},
  {Desmars}, {Vieira-Martins}, {Kilic}, {Gomes J{\'u}nior}, {Camargo},
  {Emilio}, {Vara-Lubiano}, {Kretlow}, {Albert}, {Alcock}, {Ball}, {Bender},
  {Buie}, {Butterfield}, {Camarca}, {Castro-Chac{\'o}n}, {Dunford}, {Fisher},
  {Gamble}, {Geary}, {Gnilka}, {Green}, {Hartman}, {Huang}, {Januszewski},
  {Johnston}, {Kagitani}, {Kamin}, {Kavelaars}, {Keller}, {de Kleer}, {Lehner},
  {Luken}, {Marchis}, {Marlin}, {McGregor}, {Nikitin}, {Nolthenius}, {Patrick},
  {Redfield}, {Rengstorf}, {Reyes-Ruiz}, {Seccull}, {Skrutskie}, {Smith},
  {Sproul}, {Stephens}, {Szentgyorgyi}, {S{\'a}nchez-Sanju{\'a}n}, {Tatsumi},
  {Verbiscer}, {Wang}, {Yoshida}, {Young}, \&
  {Zhang}}]{Pereiraetal2023A&A...673L...4P}
{Pereira}, C.~L., {Sicardy}, B., {Morgado}, B.~E., {et~al.} 2023, \aap, 673, L4

\bibitem[{{Pham} {et~al.}(2024){Pham}, {Rein}, \&
  {Spiegel}}]{Phametal2024OJAp....7E...1P}
{Pham}, D., {Rein}, H., \& {Spiegel}, D.~S. 2024, The Open Journal of
  Astrophysics, 7, 1

\bibitem[{Philpotts \& Ague(2009)}]{Philpotts2009}
Philpotts, A.~R. \& Ague, J.~J. 2009, Principles of Igneous and Metamorphic
  Petrology, 2nd edn. (Cambridge, UK: Cambridge University Press)

\bibitem[{{Radzievskii} \& {Artem'ev}(1962)}]{Radzievskii1962SvA.....5..758R}
{Radzievskii}, V.~V. \& {Artem'ev}, A.~V. 1962, \sovast, 5, 758

\bibitem[{{Reg{\'a}ly} {et~al.}(2018){Reg{\'a}ly}, {Dencs}, {Mo{\'o}r}, \&
  {Kov{\'a}cs}}]{Regalyetal2018MNRAS.473.3547R}
{Reg{\'a}ly}, Z., {Dencs}, Z., {Mo{\'o}r}, A., \& {Kov{\'a}cs}, T. 2018,
  \mnras, 473, 3547

\bibitem[{{Ribeiro} {et~al.}(2023){Ribeiro}, {Winter}, {Madeira}, \& {Giuliatti
  Winter}}]{Ribeiro2023MNRAS.525...44R}
{Ribeiro}, T., {Winter}, O.~C., {Madeira}, G., \& {Giuliatti Winter}, S.~M.
  2023, \mnras, 525, 44

\bibitem[{Rubincam(2000)}]{Rubincam2000}
Rubincam, D.~P. 2000, Icarus, 148, 2

\bibitem[{{Rubincam}(2006)}]{Rubincam2006Icar..184..532R}
{Rubincam}, D.~P. 2006, \icarus, 184, 532

\bibitem[{{Sachse}(2018)}]{Sachse2018Icar..303..166S}
{Sachse}, M. 2018, \icarus, 303, 166

\bibitem[{{Sfair} \& {Giuliatti Winter}(2009)}]{Sfair2009A&A...505..845S}
{Sfair}, R. \& {Giuliatti Winter}, S.~M. 2009, \aap, 505, 845

\bibitem[{Shapiro(1963)}]{Shapiro1963}
Shapiro, I.~I. 1963, in Dynamics of Satellites / Dynamique des Satellites, ed.
  M.~Roy (Berlin, Heidelberg: Springer Berlin Heidelberg), 257--312

\bibitem[{{Showalter} \& {Lissauer}(2006)}]{Showalter2006Sci...311..973S}
{Showalter}, M.~R. \& {Lissauer}, J.~J. 2006, Science, 311, 973

\bibitem[{{Showalter} {et~al.}(2006){Showalter}, {Lissauer}, \& {de
  Pater}}]{Showalter2006DPS....38.3808S}
{Showalter}, M.~R., {Lissauer}, J.~J., \& {de Pater}, I. 2006, in AAS/Division
  for Planetary Sciences Meeting Abstracts, Vol.~38, AAS/Division for Planetary
  Sciences Meeting Abstracts \#38, 38.08

\bibitem[{{Sicardy}(2020)}]{Sicardy2020AJ....159..102S}
{Sicardy}, B. 2020, \aj, 159, 102

\bibitem[{{Sicardy} {et~al.}(2025){Sicardy}, {El Moutamid}, {Renner}, {Sfair},
  \& {Souami}}]{Sicardy2025RSPTA.38340193S}
{Sicardy}, B., {El Moutamid}, M., {Renner}, S., {Sfair}, R., \& {Souami}, D.
  2025, Philosophical Transactions of the Royal Society of London Series A,
  383, 20240193

\bibitem[{{Sicardy} {et~al.}(2019){Sicardy}, {Leiva}, {Renner}, {Roques}, {El
  Moutamid}, {Santos-Sanz}, \& {Desmars}}]{Sicardy2019NatAs...3..146S}
{Sicardy}, B., {Leiva}, R., {Renner}, S., {et~al.} 2019, Nature Astronomy, 3,
  146

\bibitem[{{Sicardy} {et~al.}(2020){Sicardy}, {Renner}, {Leiva}, {Roques}, {El
  Moutamid}, {Santos-Sanz}, \& {Desmars}}]{Sicardyetal2020tnss.book..249S}
{Sicardy}, B., {Renner}, S., {Leiva}, R., {et~al.} 2020, in The Trans-Neptunian
  Solar System, ed. D.~{Prialnik}, M.~A. {Barucci}, \& L.~{Young}, 249--269

\bibitem[{{Sickafoose} \& {Lewis}(2024)}]{Sickafoose2024PSJ.....5...32S}
{Sickafoose}, A.~A. \& {Lewis}, M.~C. 2024, PSJ, 5, 32

\bibitem[{{Smith} {et~al.}(1989){Smith}, {Soderblom}, {Banfield}, {Barnet},
  {Basilevksy}, {Beebe}, {Bollinger}, {Boyce}, {Brahic}, {Briggs}, {Brown},
  {Chyba}, {Collins}, {Colvin}, {Cook}, {Crisp}, {Croft}, {Cruikshank},
  {Cuzzi}, {Danielson}, {Davies}, {de Jong}, {Dones}, {Godfrey}, {Goguen},
  {Grenier}, {Haemmerle}, {Hammel}, {Hansen}, {Helfenstein}, {Howell}, {Hunt},
  {Ingersoll}, {Johnson}, {Kargel}, {Kirk}, {Kuehn}, {Limaye}, {Masursky},
  {McEwen}, {Morrison}, {Owen}, {Owen}, {Pollack}, {Porco}, {Rages}, {Rogers},
  {Rudy}, {Sagan}, {Schwartz}, {Shoemaker}, {Showalter}, {Sicardy},
  {Simonelli}, {Spencer}, {Sromovsky}, {Stoker}, {Strom}, {Suomi}, {Synott},
  {Terrile}, {Thomas}, {Thompson}, {Verbiscer}, \&
  {Veverka}}]{Smith1989Sci...246.1422S}
{Smith}, B.~A., {Soderblom}, L.~A., {Banfield}, D., {et~al.} 1989, Science,
  246, 1422

\bibitem[{{Smith} {et~al.}(1986){Smith}, {Soderblom}, {Beebe}, {Bliss},
  {Boyce}, {Brahic}, {Briggs}, {Brown}, {Collins}, {Cook}, {Croft}, {Cuzzi},
  {Danielson}, {Davies}, {Dowling}, {Godfrey}, {Hansen}, {Harris}, {Hunt},
  {Ingersoll}, {Johnson}, {Krauss}, {Masursky}, {Morrison}, {Owen}, {Plescia},
  {Pollack}, {Porco}, {Rages}, {Sagan}, {Shoemaker}, {Sromovsky}, {Stoker},
  {Strom}, {Suomi}, {Synnott}, {Terrile}, {Thomas}, {Thompson}, \&
  {Veverka}}]{Smith1986Sci...233...43S}
{Smith}, B.~A., {Soderblom}, L.~A., {Beebe}, R., {et~al.} 1986, Science, 233,
  43

\bibitem[{{Sucerquia} {et~al.}(2017){Sucerquia}, {Alvarado-Montes},
  {Ram{\'\i}rez}, \& {Zuluaga}}]{Sucerquia2017MNRAS.472L.120S}
{Sucerquia}, M., {Alvarado-Montes}, J.~A., {Ram{\'\i}rez}, V., \& {Zuluaga},
  J.~I. 2017, \mnras, 472, L120

\bibitem[{{Sucerquia} {et~al.}(2024){Sucerquia}, {Alvarado-Montes}, {Zuluaga},
  {Cuello}, {Cuadra}, \& {Montesinos}}]{Sucerquia2024A&A...691A..74S}
{Sucerquia}, M., {Alvarado-Montes}, J.~A., {Zuluaga}, J.~I., {et~al.} 2024,
  \aap, 691, A74

\bibitem[{{Sumida} {et~al.}(2020){Sumida}, {Ishizawa}, {Hosono}, \&
  {Sasaki}}]{Sumida2020ApJ...897...21S}
{Sumida}, I., {Ishizawa}, Y., {Hosono}, N., \& {Sasaki}, T. 2020, \apj, 897, 21

\bibitem[{{Thomson} {et~al.}(2005){Thomson}, {Wong}, {Marouf}, {Rappaport},
  {French}, \& {McGhee}}]{Thomson2005AGUFM.P33B0247T}
{Thomson}, F., {Wong}, K., {Marouf}, E., {et~al.} 2005, in AGU Fall Meeting
  Abstracts, Vol. 2005, P33B--0247

\bibitem[{{Vokrouhlick{\'y}} {et~al.}(2007){Vokrouhlick{\'y}}, {Nesvorn{\'y}},
  {Dones}, \& {Bottke}}]{Vokrouhlicky2007A&A...471..717V}
{Vokrouhlick{\'y}}, D., {Nesvorn{\'y}}, D., {Dones}, L., \& {Bottke}, W.~F.
  2007, \aap, 471, 717

\bibitem[{{Warren} \& {Brandt}(2008)}]{Warren2008JGRD..11314220W}
{Warren}, S.~G. \& {Brandt}, R.~E. 2008, Journal of Geophysical Research
  (Atmospheres), 113, D14220

\bibitem[{Wiscombe(1980)}]{Wiscombe1980}
Wiscombe, W.~J. 1980, Appl. Opt., 19, 1505

\bibitem[{{Wyatt} {et~al.}(1999){Wyatt}, {Dermott}, {Telesco}, {Fisher},
  {Grogan}, {Holmes}, \& {Pi{\~n}a}}]{Wyatt1999ApJ...527..918W}
{Wyatt}, M.~C., {Dermott}, S.~F., {Telesco}, C.~M., {et~al.} 1999, \apj, 527,
  918

\bibitem[{{Wyatt} \& {Whipple}(1950)}]{Wyatt1950ApJ...111..134W}
{Wyatt}, S.~P. \& {Whipple}, F.~L. 1950, \apj, 111, 134

\bibitem[{{Zubko} {et~al.}(1996){Zubko}, {Mennella}, {Colangeli}, \&
  {Bussoletti}}]{Zubko1996MNRAS.282.1321Z}
{Zubko}, V.~G., {Mennella}, V., {Colangeli}, L., \& {Bussoletti}, E. 1996,
  \mnras, 282, 1321

\end{thebibliography}

\begin{appendix}

\section{Shadow detection algorithm }
\label{apx:shadow}

The shadow detection algorithm is designed to determine whether a given particle resides within the shadow cast by the planetary body.
This calculation is critical for modeling the effect of radiation pressure.
The method considers the relative positions of the particle and the planetary body, as well as the effective size of the planet, parameterized by its radius.

\begin{figure}
    \centering
    \includegraphics[width=1\linewidth]{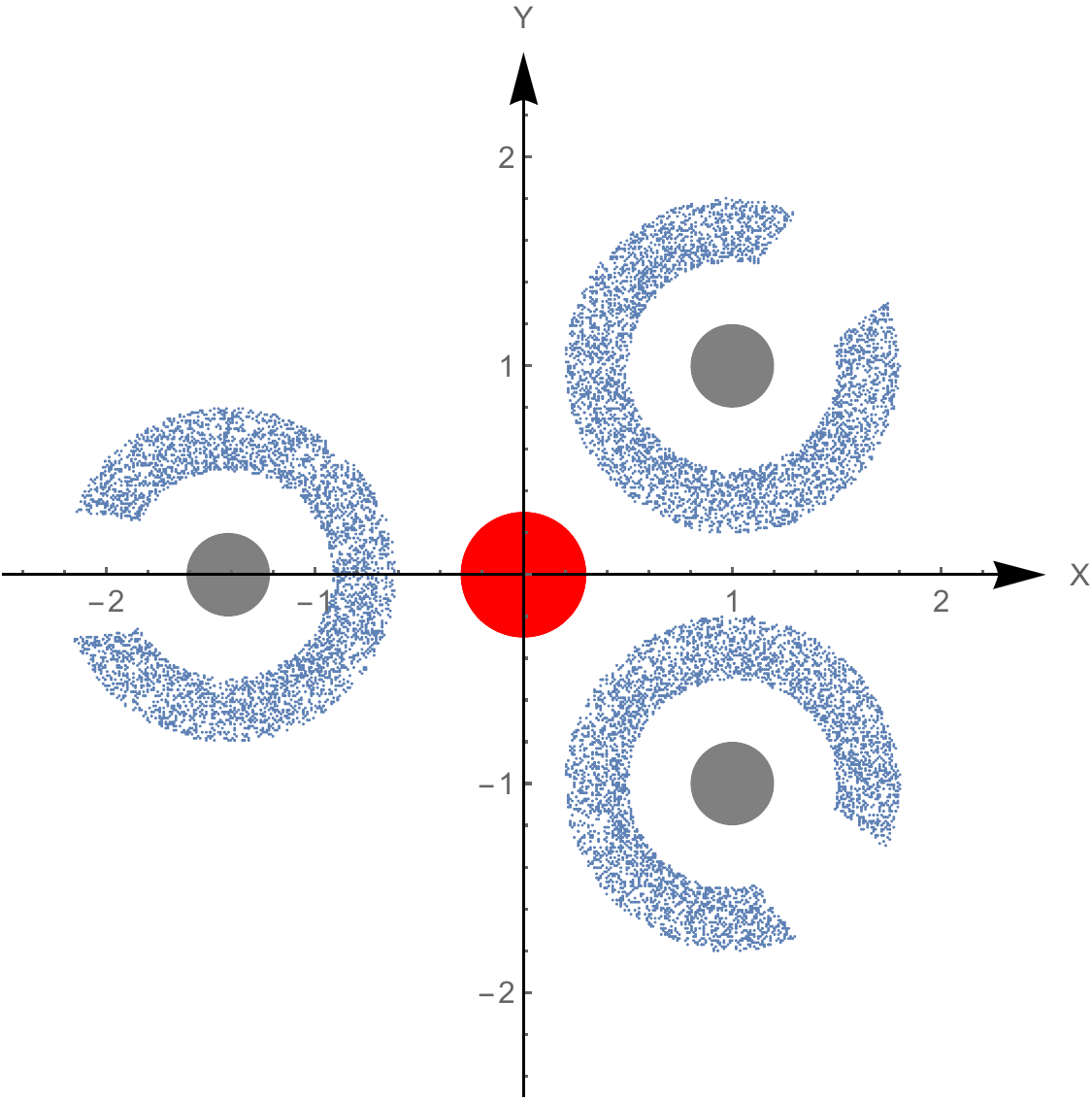}
    \caption{
    Rings in three arbitrary positions distributed in the XY plane are subject to the effects of radiation from the central object, resulting in the casting of shadows.
    Only particles that are not in the shadow of the planetary body are displayed. Note that the figure is not to scale.
    }
    \label{fig:sahadow}
\end{figure}

Two geometric conditions are evaluated to assess shadow casting.
The first condition ensures that the particle is closer to the source of the radiation than the planet, which is expressed as $R_i^2 < R_{\mathrm{p}}^2$, where $R_i$ is the distance of the ring particle and $R_{\mathrm{p}}$ that of the planetary body form the source of radiation.
This ensures that the particle lies outside the potential shadow region.
The second condition determines whether the particle falls within the shadow cone formed by the planet.
Specifically, the algorithm checks whether
\begin{equation}
\frac{(\mathbf{R}_{\mathrm{p}} \cdot \mathbf{R}_i)^2}{R_{\mathrm{p}}^2 R_i^2} < \frac{R_{\mathrm{p}}^2}{R_{\mathrm{p}}^2 + \left(D_{\mathrm{p}}/2\right)^2},
\end{equation}
where $D_\mathrm{p}$ is the diameter and $\mathbf{R}_{\mathrm{p}}$ the position vector of the planetary body, while $\mathbf{R}_i$ is the position vector of the particle.
This formulation ensures that the particle lies outside the projection of the planet's shadow.
The two conditions are combined using a logical OR operation, resulting in the final determination of shadow casting:
\begin{equation}
\gamma_\mathrm{shadow} = (R_i^2 < R_{\mathrm{p}}^2) \lor \left(\frac{(\mathbf{R}_{\mathrm{p}} \cdot \mathbf{R}_i)^2}{R_{\mathrm{p}}^2 R_i^2} < \frac{R_{\mathrm{p}}^2}{R_{\mathrm{p}}^2 + \left(D_{\mathrm{p}}/2\right)^2}\right).
\end{equation}
If the above combined condition evaluates to true, the particle is considered to be in shadow; otherwise, it is illuminated.
In Fig.~\ref{fig:sahadow}, the shadow determination process is demonstrated through three distinct configurations of the planetary body and its ring.
 
The algorithm should be evaluated for each particle at every timestep. 
Therefore, avoiding costly trigonometric computations is optimized by relying on algebraic expressions involving squared quantities.
Such efficiency is essential for simulations involving large numbers of particles, which is common in astrophysical studies of planetary systems, rings, and accretion disks \citep{lissauer1993planetary}.

\section{Particle size modeled}
\label{apx:beta-Mie}

In the expression of the relative radiation pressure force, $\beta$, given by Eq.~(\ref{eq:beta-Qprs}), the parameter $\left<Q_{\text{pr}}(s)\right>$ quantifies the efficiency of momentum transfer from solar radiation to the particle with a radius $s$.
$\left<Q_{\mathrm{pr}}(s)\right>$ is calculated by integrating the wavelength-dependent extinction and scattering efficiencies over the solar spectrum, weighted by the Sun’s emission flux \citep{Burns1979,bohren1983}.
Specifically,
\begin{equation}
\left<Q_{\mathrm{pr}}(s)\right> = \frac{\int Q_{\mathrm{pr}}(\lambda, s) F_{\odot}(\lambda) \, d\lambda}{\int F_{\odot}(\lambda) \, d\lambda}.
\end{equation}
The solar flux at a given wavelength $\lambda$ is
\begin{equation}
    F_{\odot}(\lambda)\sim\frac{2 h \nu^3}{c^2}\frac{1}{e^{h \nu / (k T_\odot)}-1},
\end{equation}
where $h$ and $k$ are the Planck and the Boltzmann constants, respectively, while $T_\odot=5800$~K is the effective solar temperature.
$Q_{\mathrm{pr}}(\lambda, s)$ includes contributions from absorption and scattering, with 
\begin{equation}
Q_{\text{pr}}(\lambda,s) = Q_{\text{abs}}(\lambda,s)  + Q_{\text{sca}}(\lambda,s) (1 - g),
\end{equation}
where $Q_{\text{abs}}(\lambda,s) $ is the absorption efficiency, $Q_{\text{sca}}(\lambda,s) $ is the scattering efficiency, and $g$ is the asymmetry parameter describing the scattering anisotropy phase function for a given wavelength and particle size.

The calculation of $Q_{\text{pr}}(\lambda,s) $ often employs Mie scattering theory, which provides an exact solution for the scattering and absorption of electromagnetic waves by homogeneous spherical particles.
The Mie scattering coefficients $a_n$ and $b_n$, derived from boundary conditions for electromagnetic waves, are used to compute the efficiencies:
\begin{equation}
Q_{\text{abs}}(\lambda,s)  + Q_{\text{sca}}(\lambda,s)  = \frac{2}{s^2} \sum_{n=1}^\infty (2n + 1)\Re(a_n + b_n),
\end{equation}
\begin{equation}
Q_{\text{sca}}(\lambda,s)  = \frac{2}{s^2} \sum_{n=1}^\infty (2n + 1)\left(|a_n|^2 + |b_n|^2\right),
\end{equation}
where the terms $|a_n|^2$ and $|b_n|^2$ are the magnitudes of the Mie coefficients.
Mie scattering theory calculates $Q_{\text{abs}}(\lambda,s) $ and $Q_{\text{sca}}(\lambda,s) $ based on the particle size parameter $x = 2 \pi s / \lambda$ and the complex refractive index of the material $m = n + ik$ \citep{Wiscombe1980}.
The implementation of Mie scattering theory for the calculation of $Q_{\text{sca}}(\lambda,s)$, typically involves the use of numerical solutions due to the inherent complexity of the infinite series associated with this calculation.

For various materials such as water ice, graphite, iron, and basalt, the determination of $Q_{\mathrm{pr}}(\lambda,s)$ requires accurate optical constants, often sourced from laboratory measurements or established databases (see, e.g., \citealp{Palik1991}).
For instance, water ice typically exhibits $Q_{\mathrm{pr}}(\lambda,s)$ values around unity for micron-sized grains due to its modest absorption and scattering in the visible spectrum.
In contrast, graphite and metallic iron particles may display higher $Q_{\mathrm{pr}}(\lambda,s)$ values owing to their strong absorption and complex scattering behaviors.
Basaltic particles, resembling silicate dust, generally show $Q_{\mathrm{pr}}(\lambda,s)$ values ranging from 0.5 to 1.5, influenced by their mixed mineral compositions and varying optical properties.

\begin{figure}
    \centering
    \includegraphics[width=1\linewidth]{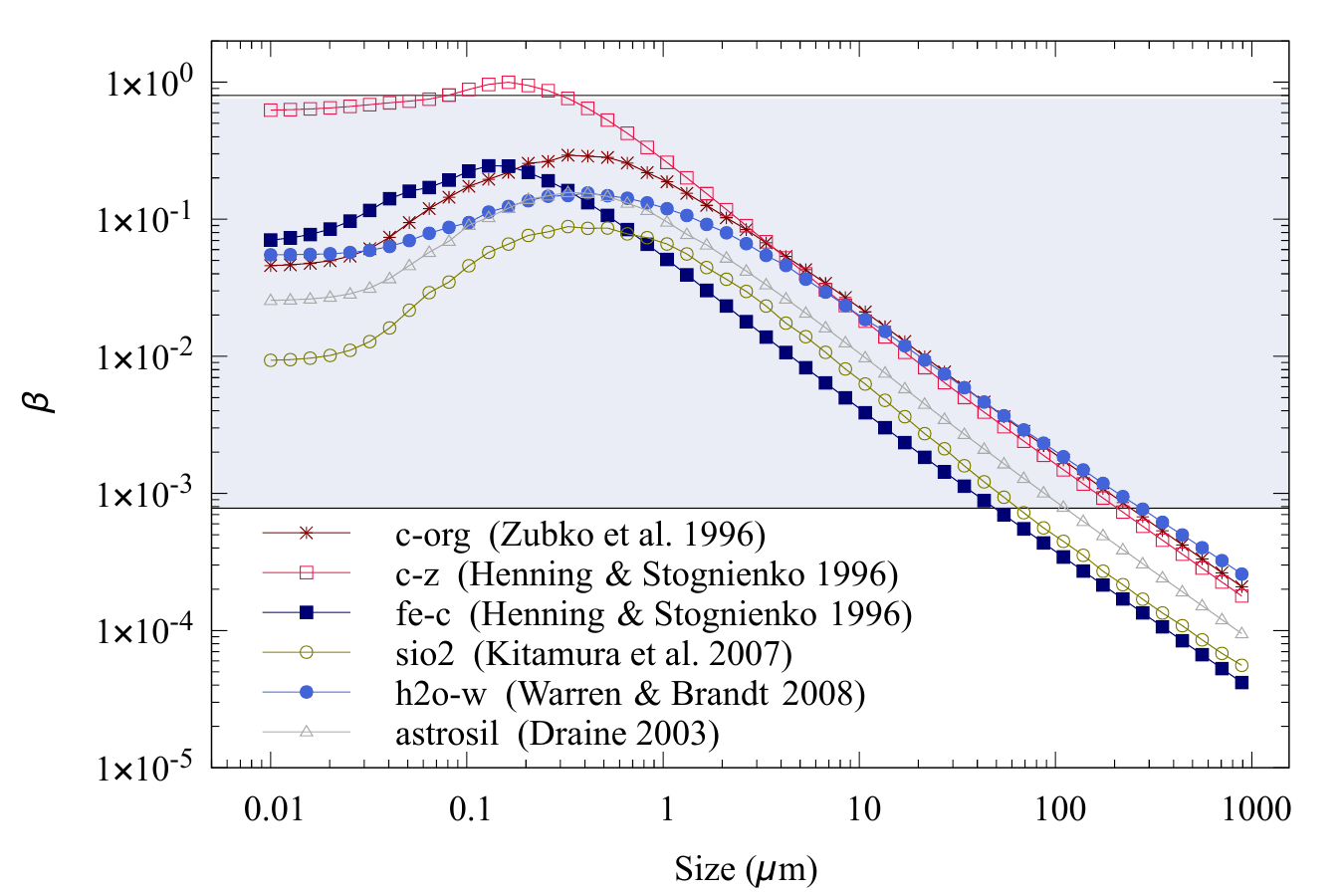}
    \caption{Normalized radiation pressure $\beta$, as a function of grain size $s$. 
    Six different grain compositions are presented. Calculations assume homogeneous spherical grains in Mie theory.}
    \label{fig:beta-size}
\end{figure}

For this study, we used the software package \verb|optool| to calculate $Q_{\text{pr}}(\lambda,s)$ provided by \citet{Dominik2021ascl.soft04010D}.
We assume classical Mie scattering theory, meaning that the grains are assumed to be homogeneous spherical particles.
$\beta$ is calculated for the following compositions: amorphous organic carbon, ({\small c-org} \citealp{Henning1996A&A...311..291H}); amorphous carbon, ({\small c-z} \citealp{Zubko1996MNRAS.282.1321Z}); metallic iron, ({\small fe-c} \citealp{Henning1996A&A...311..291H}); amorphous quartz representing basalt\footnote{Basalt generally has a composition of 45–52\% $\mathrm{SiO_2}$, 2–5\% total alkalis, 0.5–2.0\% $\mathrm{TiO_2}$, 5–14\% FeO and 14\% or more $\mathrm{Al_2O_3}$. Contents of CaO are commonly near 10\%, those of MgO commonly in the range of 5-12\% \citep{Philpotts2009,Irvine1971CaJES...8..523I}} ({\small sio2} \citealp{Kitamura2007ApOpt..46.8118K}); water ice, ({\small h2o-w} \citealp{Warren2008JGRD..11314220W}); and astronomical silicate, ({\small astrosil} \citealp{Draine2003ApJ...598.1017D}).
Figure~\ref{fig:beta-size} shows $\beta$ as a function of grain size for the six different compositions.
As demonstrated in the figure, the range of $7.8\times10^{-4}\leq\beta\leq0.8$ corresponds to grain sizes ranging from $0.1~\mu$m to $0.5$~mm.

\end{appendix}

\end{document}